\documentclass[aps,twocolumn]{revtex4}

\usepackage{amsmath}
\usepackage{graphicx}
\usepackage{amssymb}
\usepackage{subfigure}

\newcommand{\beq}{\begin{equation}}
\newcommand{\eeq}{\end{equation}}

\newcommand{\bfx}{({\bf x})}
\newcommand{\ud}{\mathrm{d}}

\newcommand{\si}{\sigma}
\newcommand{\om}{\omega}
\newcommand{\De}{\Delta}
\newcommand{\de}{\delta}

\begin{document}
\title[]{Ultracold atoms in optical lattices generated by quantized light fields}


\author{Christoph Maschler$^{1}$}
\author{Igor B. Mekhov$^{1,2}$}
\author{Helmut Ritsch$^{1}$}%

\affiliation{$^1$Institut f\"ur theoretische Physik, Universit\"at Innsbruck, Technikerstr.~25, A-6020 Innsbruck,Austria\\
$^2$St. Petersburg State University, Faculty of Physics, St. Petersburg, Russia}


\begin{abstract}
We study an ultracold gas of neutral atoms subject to the periodic optical potential generated by a high-$Q$ cavity mode.
In the limit of very low temperatures, cavity field and atomic dynamics require a quantum description. Starting from a cavity QED single atom Hamiltonian we use different routes to derive approximative multiparticle Hamiltonians in Bose-Hubbard form with rescaled or even dynamical parameters. In the limit of large enough cavity damping the different models agree. Compared to free space optical lattices, quantum uncertainties of the potential and the possibility of atom-field entanglement lead to modified phase transition characteristics, the appearance of new phases or even quantum superpositions of different phases. Using a corresponding effective master equation, which can be numerically solved for few particles, we can study time evolution including dissipation. As an example we exhibit the microscopic processes behind the transition dynamics from a Mott insulator like state to a self-ordered superradiant state of the atoms, which appears as steady state for transverse atomic pumping.
\end{abstract}


\maketitle

\section{Introduction}

Laser light, far red detuned from an atomic resonance, is nowadays a
standard tool in experimental quantum optics to create tunable
optical potentials~\cite{lasercool} which can be loaded with ultracold atoms
to provide for a plethora of possibilities  to study quantum properties of
many-body strongly correlated systems~\cite{Bloch05}. The high level of microscopic understanding
and extensive control of the light fields and atoms allow to implement genuine models like e.g. the Bose-Hubbard (BH) model~\cite{Jaksch98,Zwerger03}.
Initially originating from condensed matter physics~\cite{Fisher89} it has been used to study the Mott insulator to
superfluid phase transition~\cite{Greiner02} in detail and in real time.
Adjusting several of the lattice parameters as the intensity and the configuration of the
lattice lasers provides a versatile toolbox of techniques to control
the dynamics of the atoms in the lattice~\cite{Jaksch05}. Moreover,
the collisional properties of the certain types of atoms can be
tailored by means of magnetic~\cite{Inouye98} or optical~\cite{Theis04} Feshbach resonances.
Using extra confinement it was even possible to observe the Mott insulator to superfluid transition in 1D~\cite{Stoeferle04,Koehl05}
and 2D~\cite{Spielman07}, followed by other spectacular demonstrations of condensed
matter physics phenomena as the realization of a Tonks gas in 1D~\cite{Paredes04,Kinoshita04} and the
Berezinskii-Kosterlitz-Thouless phase transition in 2D~\cite{hadzibabic06}. Theoretically many more proposals
to apply these methods to spin systems and investigate further fascinating properties of strongly correlated systems
were put forward (see~\cite{lewenstein_review} for a review).

In all of these approaches, the light fields were approximated by classical, externally prescribed fields
independent of the atoms. This requires intense light, far detuned from any atomic transition. Of course this assumption holds no
longer if the light, which generates the optical lattice, is enhanced by an optical resonator. In this case - given a sufficient atom number $N$ and atom-field coupling $g$ -
the field itself becomes a dynamical quantity~\cite{Domokos03} depending on the atomic distribution. As all atoms are coupled to the
same field modes, this immediately introduces substantial long range interactions, which cannot be ignored as in free space.
In specially designed cases this force induces coherently driven atoms to self-organize in regular patterns as predicted in Ref.~\cite{domokos02,Zippilli04a} and subsequently
experimentally verified~\cite{black03}.

In addition, in a high-$Q$ optical resonator relatively low photon
numbers are sufficient to provide strong forces. This was
demonstrated by trapping an atom in the field of just a single
photon~\cite{Pinkse00,Hood00}. Hence the inevitable photon number
fluctuations induced by cavity damping generate force fluctuations
on the atoms causing diffusion. At the same time as cavity photon
loss constitutes a dissipation channel, it can also carry out energy
and entropy of the system. This opens possibilities for cooling of
atomic motion ~\cite{Horak97,Vuletic01,Domokos04,Zippilli05}, as
demonstrated by beautiful experiments in the group of
Rempe~\cite{maunz04,Nussmann05}. Since this cooling mechanism does
not require the existence of closed optical cycles it could even be
used for qubits~\cite{Griessner04} or to damp quantum oscillations
or phase fluctuations of a BEC coupled to a cavity
field~\cite{Horak00,Jaksch01}.

For low photon numbers the quantum properties of the light field get important as well and the atoms are now moving in different quantized potentials determined by
the cavity photon number. Quantum mechanics of course allows for superpositions of photon numbers invoking superpositions of different optical potentials for the atoms.
First simplified models to describe this new physics were recently proposed by us~\cite{maschler05} and in parallel by other authors~\cite{lewenstein06}.
As the intracavity field itself depends on the atomic state (phase), different atomic quantum states are correlated with different states of the lattice
field with differing photon number distributions. In this way quantum mechanics allows for the creation of very exotic atom-field states, like a superposition of a Mott-insulator and superfluid
phase, each thereof correlated with a different photon number. Some quite exotic looking phase diagrams for this system were already discussed in Ref.~\cite{lewenstein06}. Without resorting to the full complex dynamics of the system,
the quantum correlations between the field and the atomic wavefunctions open the possibility of non-destructively probing the atomic state by weak scattering of coherent light into the cavity mode~\cite{mekhov1} and carefully analyzing its properties~\cite{mekhov2}.

It is quite astonishing, that experimental progress in the recent
years has made such systems experimentally accessible and at present
already several experimental groups succeeded in loading a BEC into
a high-$Q$ optical
cavity~\cite{Esslinger,Reichel07,Brennecke07,Slama07,Gupta07}. A
reliable analysis of these experiments has made more thorough
theoretical studies of such systems mandatory.

In this work we concentrate on the study of an ultracold gas in
optical lattices including the quantum nature of the lattice
potential generated from a cavity field. This extends and
substantiates previous studies and predictions on such a system by
us~\cite{maschler05} as well as other authors~\cite{lewenstein06}.
Here we limit ourselves to the case of a high-$Q$ cavity which
strongly enhances a field sufficiently red detuned from any atomic
transition to induce an optical potential without significant
spontaneous emission. In particular we address two different
geometric setups, where either the cavity mode is directly driven
through one mirror, or the atoms are coherently excited by a
transverse laser and scatter light into the cavity mode. The cavity
potential can also be additionally enhanced by some extra
conservative potential applied at a different
frequency~\cite{Mckeever03,Puppe07}. These two generic cases leads
to quite different physical behavior and allow to discuss several
important aspects of the underlying physics.

This paper is organized as follows. Sec.~\ref{sec:model} is devoted to a systematic presentation of our model and
various simplifying approximations as adiabatic elimination of the excited states of the atoms and subsequent
formulation of an effective multi-particle Hamiltonian in second quantized form.
In section~\ref{sec:cav_pump}, we specialize on the simplest generic case of a
coherently driven cavity and approximate the corresponding Hamiltonian
by adiabatic elimination of the cavity field. We investigate the properties
thereof, corresponding to the influence of the cavity on the
Mott-insulator to superfluid quantum phase-transition and identify
the regime of validity for the elimination of the cavity field.
Finally, we compare these results with the dynamics of the full
master equation. In Sec.~\ref{sec:atom_pump} we study the more complex case of atoms coherently
driven by a laser field transversal to the cavity axis, where it is much harder to find valid analytical simplifications and one has to resort to numerical studies of few particle dynamics.
Finally, we conclude in Sec.~\ref{sec:conclusions}.

\section{Model}\label{sec:model}
\begin{figure}[htp]
    \begin{center}
    \includegraphics[width=1\hsize]{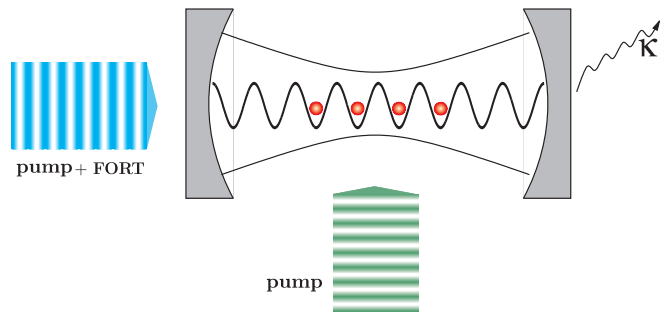}
      \caption{\label{cavity}(color online). Scheme of atoms inside an optical cavity, driven by two external pumping lasers.
      An additional conservative lattice potential, independent of the intracavity field, is realized by a far off-resonant dipole trap (FORT).}
    \nobreak\medskip
   \end{center}
  \end{figure}

We start with $N$ two-level atoms with mass $m$ and transition
frequency $\omega_{eg}$ strongly interacting with a single standing
wave cavity mode of frequency $\omega_c$. We also consider coherent
driving of the atoms at frequency $\omega_p$ and with maximal
coupling strength $h_0$ and of the cavity with amplitude $\eta$ (see
Fig.~\ref{cavity}). Note that in the specific examples later we will
consider only one pump laser beam at a time.

Using the rotating-wave and electric-dipole approximation, we can
describe a single atom of this system by the Jaynes-Cummings
Hamiltonian~\cite{Jaynes63} \beq H^{(1)} = H_{A}^{(1)} + H_R^{(1)} +
H_{Int}^{(1)}.\label{eq:spHam}\eeq
Explicitly the different Hamiltonians for the atoms, the field mode and the interaction read:
 \begin{subequations}\label{eq:singlepartham1}
\begin{align}
  H^{(1)}_A &  = \frac{\hat{{\bf p}}^2}{2m} + V_e\bfx\si^+\si^- + V_g\bfx\si^-\si^+ +\hbar\om_{eg}\si^+\si^-\nonumber\\& - i\hbar h\bfx\left(\si^+e^{-i\om_pt}-\si^-e^{i\om_pt}\right),\\
H_R^{(1)} & = \hbar\om_c a^\dag a - i\hbar \eta \left( a e^{i\om_p t}-a^\dag e^{-i\om_p t}\right),\\
H_{Int}^{(1)} & =  - i\hbar g\bfx\left(\si^+ a-\si^-a^\dag\right).
\end{align}
\end{subequations}
Here $h\bfx$ denotes the mode-function of the transverse pump field,
$g\bfx$ denotes the cavity mode function and the field operator $a$
describes the annihilation of a cavity photon with frequency
$\om_c$.  $V_e\bfx$ and $V_g\bfx$ are external trapping potentials
for the atom in the excited and the ground state, respectively. In
order to change to slowly varying variables we apply a unitary
transformation with operator $U(t) =
\exp[i\om_pt\left(\si^+\si^-+a^\dag a\right)]$, such that we end up
with the following single-particle Hamiltonian, using the same
symbols for the transformed quantities:
\begin{subequations}\label{eq:spHam2}
\begin{align}
  H^{(1)}_{A} &  = \frac{\hat{{\bf p}}^2}{2m} + V_e\bfx\si^+\si^- + V_g\bfx\si^-\si^+ - \hbar\De_a\si^+\si^-\nonumber\\&
  - i\hbar h\bfx\left(\si^+-\si^-\right),\\
H_{R}^{(1)} & = -\hbar\De_c a^\dag a - i\hbar \eta \left( a -a^\dag\right),\\
H_{Int}^{(1)} & =  - i\hbar g\bfx\left(\si^+ a-\si^-a^\dag\right),
\end{align}
\end{subequations}
where $\De_c=\om_p-\om_c$, $\De_a=\om_p-\om_{eg}$ denotes the detunings of
the cavity and the atomic transition frequency from the pumping
field frequency. In order to describe the situation for $N$ atoms,
we use the single-particle Hamiltonian of Eq.~\eqref{eq:spHam}
and~\eqref{eq:spHam2} in second quantization formalism\cite{Galindo}, i.e.,
\beq
H=H_{A}+H_{R}+H_{A-R}+H_{A-P}+H_{A-A}.
\eeq
The terms in this expression correspond to the single particle terms in~\eqref{eq:singlepartham1} and~\eqref{eq:spHam2}.
Hence, $H_{A}$ and $H_{R}$ model the free evolution of the atomic and the field variables, respectively. They read as:
\begin{multline}
H_{A}=\int\ud^3 {\bf x}\left[ \Psi_g^{\dag}\bfx \left(
-\frac{\hbar^2}{2m}\nabla^2+V_g\bfx\right)\right.\Psi_g\bfx\\ +
 \Psi_e^{\dag}\bfx \left.\left( -\frac{\hbar^2}{2m}\nabla^2-\hbar\De_a+V_e\bfx\right)\Psi_e\bfx\right],
\end{multline}
where $\Psi_g\bfx$ and $\Psi_e\bfx$ denotes the atomic field
operators for annihilating an atom at position ${\bf x}$ in the
ground state and the excited state, respectively. They obey the
usual bosonic commutation relations
\begin{subequations}
\begin{align}
\left[ \Psi_f\bfx,\Psi_{f^\prime}^\dag({\bf x}^\prime)\right] & =
\de^3\left({\bf x}-{\bf x}^\prime\right)\de_{f,f^\prime}\\ \Big{[}
\Psi_f\bfx,\Psi_{f^\prime}({\bf x}^\prime)\Big{]} & =\left[
\Psi_f^\dag\bfx,\Psi_{f^\prime}^\dag({\bf x}^\prime)\right]=0,
\end{align}
\end{subequations}
for $f,f^\prime\in\{e,g\}$. The field operator remains unchanged,
i.e., $H_{R}=-\hbar\De_c a^\dag a - i\hbar \eta \left( a
-a^\dag\right)$.
 The two-body interaction is modeled by a short-range pseudopotential~\cite{huang}, characterized by the s-wave scattering length $a_s$, leading to a Hamiltonian
\beq H_{A-A}=\frac{U}{2}\int\ud^3 {\bf
x}\Psi^\dag_g\bfx\Psi^\dag_g\bfx\Psi_g\bfx\Psi_g\bfx, \eeq where
$U=4\pi a_s\hbar^2/m$. The coupling of the cavity field with the
atoms inside the cavity is given by \beq H_{A-R}=-i\hbar\int\ud^3
{\bf x}\Psi_g^\dag\bfx g\bfx a^\dagger \Psi_e\bfx + \mathrm{h.c.},
\eeq whereas the interaction with the laser beam, which coherently
drives the atoms, reads \beq H_{A-P}=-i\hbar\int\ud^3 {\bf
x}\Psi_g^\dag\bfx h\bfx  \Psi_e\bfx +  \mathrm{h.c.}. \eeq

Let us now calculate the Heisenberg equations for the various field operators, starting with the operator for the excited state, i.e.,
\begin{multline}\label{eq:psi_e}
\frac{\partial \Psi_e\bfx}{\partial t} =
i\left[\frac{\hbar}{2m}\nabla^2-\frac{V_e\bfx}{\hbar}+\De_a\right]\Psi_e\bfx\\-\left[
g\bfx a +  h\bfx\right]\Psi_g\bfx.
\end{multline}
The first term corresponds to the free evolution of the atomic state,
whereas the second term describes the absorption of a cavity
photon or a pump photon together with the annihilation of a ground state atom.
Similarly, the equation for the ground state operator reads:
\begin{multline}\label{eq:psi_g}
\frac{\partial \Psi_g\bfx}{\partial t} =
i\left[\frac{\hbar}{2m}\nabla^2-\frac{V_g\bfx}{\hbar}-\frac{U}{\hbar}\Psi^\dag_g\bfx\Psi_g\bfx\right]\Psi_g\bfx\\+\left[
g\bfx a^\dagger +  h\bfx\right]\Psi_e\bfx.
\end{multline}
Finally, the Heisenberg equation for the cavity field operator is given by:
\beq\label{eq:a}
\frac{\partial a}{\partial t} = i\De_c a + \eta + \int\ud^3 {\bf x}g\bfx\Psi_g^\dag\bfx\Psi_e\bfx.
\eeq
Again, the first term corresponds to the free field evolution,
whereas the last two terms are driving terms of the cavity field.

As we want to treat temperatures close to $T=0$ we have to avoid
heating
and ensure weak atomic excitation, where there is only negligible
spontaneous emission. In this limit we can adiabatically eliminate
the excited states from the dynamics of our system. This requires
large atom-pump detunings $\De_a$, where we also can neglect the
kinetic energy term and the trapping potential in~(\ref{eq:psi_e})
compared to $\De_a$. Necessarily, we assume that the field operators
$\Psi_g\bfx$ and $a$ vary on a much slower time scale than the
$1/\De_a$ terms, such that we obtain: \beq \Psi_e({\bf
x},t)=-\frac{i}{\De_a}\left[ h\bfx+g\bfx a(t)\right]\Psi_g({\bf
x},t). \eeq

Inserting this expression for $\Psi_e\bfx$ into~(\ref{eq:psi_g}) and~(\ref{eq:a}) leads then to:
\begin{multline}\label{eq:ad_ham}
\frac{\partial \Psi_g\bfx}{\partial t} =
i\left[\frac{\hbar}{2m}\nabla^2-\frac{V_g\bfx}{\hbar}-\frac{h^2({\bf
x})}{\De_a}-\frac{g^2({\bf x})}{\De_a}a^\dag a\right.\\\left.-\frac{
h\bfx g\bfx}{\De_a}\left( a +
a^\dag\right)-\frac{U}{\hbar}\Psi^\dag_g\bfx\Psi_g\bfx\right]\Psi_g\bfx,
\end{multline}
\begin{multline}\label{eq:ad_ham2}
\frac{\partial a}{\partial t} = i\left[\De_c
-\frac{1}{\De_a}\int\ud^3 {\bf x}g^2({\bf
x})\Psi_g^\dag\bfx\Psi_g\bfx\right] a\\
-\frac{i}{\De_a}\int\ud^3 {\bf x}h\bfx\Psi_g^\dag\bfx\Psi_g\bfx +
\eta.
\end{multline}

To discuss the underlying physics in a tractable form, the trick is now to find an effective Hamiltonian $H_{\textrm{eff}}$
which leads to the same dynamics as given by Eq.~(\ref{eq:ad_ham}) and~(\ref{eq:ad_ham2}).
Thus this Hamiltonian has to obey:
\beq
\label{eq:effHam2ndquant}i\hbar\frac{\partial \Psi_g\bfx}{\partial
t}=\left[ \Psi_g\bfx,H_{\textrm{eff}}\right]\,\,\,\textrm{and}\,\,\,
i\hbar\frac{\partial a}{\partial t} = \left[
a,H_{\textrm{eff}}\right].
\eeq
From this we can easily read off a possible effective Hamiltonian of the form:
\begin{multline}\label{ad_ham}
H_{\textrm{eff}}= \int\ud^3{\bf
x}\Psi_g^\dag\bfx\left\{-\frac{\hbar^2}{2m}\nabla^2+V_g\bfx\right.\\\left.+\frac{\hbar}{\De_a}\left[
h^2({\bf x})+g^2({\bf x})a^\dag a+ h\bfx g\bfx\left(
a+a^\dag\right)\right]\right\}\Psi_g\bfx\\+\frac{U}{2}\int\ud^3{\bf
x}\Psi_g^\dag\bfx\Psi_g^\dag\bfx\Psi_g\bfx\Psi_g\bfx\\-i\hbar\eta\left(
a-a^\dag\right)-\hbar\De_ca^\dag a.
\end{multline}
The corresponding single particle Hamiltonian, which leads to this second quantized Hamiltonian is~\cite{comment1}:
\begin{multline}\label{eq:adhamSP}
H_{\textrm{eff}}^{(1)}=\frac{{\bf
p}^2}{2m}+V_g\bfx+\frac{\hbar}{\De_a}\left[ h^2({\bf x})+g^2({\bf
x})a^\dag a\right.\\\left.+ h\bfx g\bfx\left(
a+a^\dag\right)\right]-i\hbar\eta\left(
a-a^\dag\right)-\hbar\De_ca^\dag a.
\end{multline}

This simplified effective atom-field Hamiltonian will be the basis
of our further considerations. It is, however, still much too
complex for a general solution and we will have to make further
simplifying assumptions. Hence at this point we will restrict
ourselves to 1D motion along the cavity axis. In an experimental
setup this could be actually realized by a deep radial trapping
potential, but we think that at least qualitatively the model should
also capture the essential physics if some transverse motion of the
particles was allowed. As one consequence this assumption requires a
rescaling of the effective two-body interaction
strength~\cite{olshanii98}, which enters as a free parameter in our
model anyway.

Mathematically we thus end up with a one-dimensional optical lattice, wich is partly generated by the resonator field
and superimposed onto a prescribed extra trapping potential $V_g\bfx=V_g(x)$. The
mode function of the cavity along the axis is approximated by $g\bfx=g(x)=g_0\cos(kx)$ and the transverse laser beam forms a broad standing wave $h({\bf x})=h_0\cos(k_py)$, which in our one-dimensional considerations ($y=0$) is just a constant term that we can eventually omit in~\eqref{ad_ham}.

As we consider external pumping of atoms and mode, we essentially treat an open system and we have to deal with dissipation
as well. Such dissipation processes are modeled by Liouvillean terms $\mathcal{L}$ appearing in the master equation for the atom-field density operator, i.e.,
 \begin{equation}\label{eq:liouville}
    \dot{\varrho}=\frac{1}{i\hbar}[H_{\textrm{eff}},\varrho]+\mathcal{L}\varrho.
 \end{equation}
As mentioned above, we assume large atom-pump detuning $\Delta_a$, suppressing spontaneous emission to a large extend.
However, we still have to deal with the cavity loss $\kappa$, which will thus be the dominant dissipation process.
Hence the corresponding Liouvillean using a standard quantum optics approach~\cite{Gardiner} reads:
 \begin{equation}\label{eq:liouvillian}
   \mathcal{L}\varrho=\kappa\left( 2a\varrho a^\dag - a^\dag a\varrho -\varrho a^\dag a\right).
  \end{equation}

Equivalently in the corresponding Heisenberg equation for the field operator, cavity loss leads to damping terms and fluctuations, so that it then reads:
\begin{multline}\label{eq:heisenberg}
      \dot{a} = \left\{i\left[\Delta_c-\frac{g_0^2}{\Delta_a}\int\ud x\Psi^\dag_g(x)\cos^2(kx)\Psi_g(x)\right]-\kappa \right\}a\\-i\frac{ g_0h_0}{\Delta_a}\int\ud x\Psi^\dag_g(x)\cos(kx)\Psi_g(x) + \eta + \Gamma_{in}.
 \end{multline}
Since we will be mainly interested in normally ordered quantities and assume vacuum (T=0) outside the cavity, the input noise operators $\Gamma_{in}$ will not enter in the dynamics, such that we will omit them later.

Let us now proceed and transform the Hamiltonian into a more commonly known form. Following standard procedures, one constructs maximally localized eigenfunctions at each site and expands the atomic field operator $\Psi_g(x)$ in terms of single atom Wannier functions~\cite{kittel}
\begin{equation}\label{Wann1}
    \Psi_g(x)=\sum_n\sum_k b_{n,k} w_n(x-x_k),
  \end{equation}
where $b_{n,k}$ corresponds to the annihilation of a particle in the
$n-$th energy band at site $k$. Since we assume the involved
energies to be much smaller than the excitation energies to the
second band, we are able to keep only the lowest vibrational state
in the Wannier expansion, i.e., $\Psi_g(x)=\sum_k b_k w(x-x_k)$,
where $w(x)=w_0(x)$. This yields to the following Hamiltonian:
\begin{eqnarray}
     \label{BHham} \nonumber H&=&\sum_{k,l}E_{kl}b_k^\dag b_l + \left(\hbar U_0 a^\dag a + V_{\textrm{cl}}\right)\sum_{k,l}J_{kl}b_k^\dag b_l\\\nonumber&+&\hbar\eta_{\textrm{eff}}\left( a+a^\dag\right)\sum_{k,l}\tilde{J}_{kl}b_k^\dag b_l-i\hbar\eta\left( a-a^\dag\right)\\&+&\frac{1}{2}\sum_{i,j,k,l}U_{ijkl}b_i^\dag b_j^\dag b_kb_l-\hbar\Delta_c a^\dag a,
\end{eqnarray}
where the addendum {\it eff} of the Hamiltonian is omitted. Here we
introduced an important characteristic parameter of atomic cavity
QED, namely the refractive index $U_0$ of a single atom at an
antinode, which is given by $U_0=g_0^2/\Delta_a$. It gives the
frequency shift of the cavity mode induced by a single atom at an
antinode and also corresponds to the optical lattice depth for an
atom per cavity photon~\cite{Domokos03}. Similarly, the parameter
$\eta_{\textrm{eff}}=g_0h_0/\Delta_a$ describes the position
dependent effective pump strength of the cavity mode induced by the
scattered light from a single atom at an antinode.

Note that the Wannier state expansion Eq.~\ref{Wann1} depends on the
potential depth. Thus the Wannier functions and the corresponding
matrix elements depend on the cavity field and thus in principle are
dynamic quantities. However, they keep the same functional form with
a few changing parameters, which have to be determined consistently.
This is of course consequently also true for the various coupling
parameters in the Hamiltonian. The above model thus can only be
valid as long as the single band approximation stays valid during
the system dynamics and the parameters dont change to rapidly. In
the special but rather interesting case, where the atoms are trapped
solely by the cavity field~\cite{Pinkse00,Hood00} this is not valid
for very low photon numbers. Here a single photon number jump will
induce excitation to higher bands, which induces nonlinear dynamics
beyond the single band model.

In practise this problem can be circumvented by adding an additional external trapping potential $V_g(x)$ to the model,
which guarantees a minimum potential depth even in the case of zero cavity photons. Experimentally this is feasible, for
instance, with a far detuned, off-resonant dipole trap (FORT)~\cite{grimm}, i.e., $V_g(x)=V_{\textrm{cl}}\cos^2(k_Fx)$,
where $k_F$ denotes the wave number of the FORT field. In the experimental realization, the frequency of the corresponding laser field $\om_F$ is only very few free spectral ranges separated from the main cavity frequency $\omega_c$~\cite{ye99,maunz04,sauer04}. Hence, in the vicinity of the cavity center, the coincidence of the
FORT field and the cavity field is very good, and we can replace in good agreement $\cos^2(k_Fx)$ with $\cos^2(kx)$.

Let us remark here that by including this extra potential, we can keep our model and allow for further analytical analysis of the dynamics, but we also have thrown out a great deal of interesting physics already. Actually, for very few atoms one still can solve the full Hamiltonian without the restriction to the lowest bands by quantum Monte Carlo wavefunction simulations. Some early results of such simulations can be found in Ref.~\cite{quantum_seesaw,quantum_seesaw2}.  However, this is not the subject of this work and we will proceed here with the effective lattice model under the assumption of a deep enough extra potential or strong enough cavity fields.

Note that in~\eqref{BHham}, in contrast to the case of the Bose-Hubbard model in a classical optical lattice, where the
matrix elements of the potential and kinetic energy can be merged, here two separate parts exist due to the presence of the cavity field operators in the Hamiltonian.
Explicitly they read as:
\begin{subequations}\label{mat_els}
\begin{align}
 E_{kl} = & \int\mathrm{d} x\, w(x -  x_k)\left( - \frac{\hbar^2}{2m}\nabla^2\right) w(x - x_l),\\
J_{kl} = & \int\mathrm{d} x\,w(x - x_k)\cos^2(kx)w(x - x_l),\\
\tilde{J}_{kl}  = & \int\mathrm{d} x\,w(x -  x_k)\cos(kx)w(x - x_l).
\end{align}
\end{subequations}
The on-site elements $J_{kk}$ and $E_{kk}$ are independent of the
lattice site $k$, whereas  $\tilde{J}_{kl}$ changes sign
periodically, i.e., $\tilde{J}_{kk}=-\tilde{J}_{k+1,k+1}$ due to the
$\cos$, which has twice the periodicity of the lattice. This also
accounts for $\tilde{J}_{k,k+1}=0$.  Note that the existence of this
term implies that two adjacent wells acquire different depths
forcing us to reassure that for the case of the directly pumped atom
$\eta_{\textrm{eff}}(a+a^\dagger)\cos(kx)$ is only a small
perturbation of the lattice. As the next-nearest elements are
typically two orders of magnitude smaller than the nearest-neighbor
term~\cite{Jaksch98} they can safely be neglected (tight-binding
approximation). Hence we label the site-independent on-site matrix
elements with $E_0, J_0$ and $\tilde{J}_0$, whereas $E$ and $J$ are
the site-to-site hopping elements. Furthermore, in the case of the
nonlinear interaction matrix elements,
\begin{equation}
   U_{ijkl} = g_{1D}\int\mathrm{d} x w(x-x_i)w(x-x_j)w(x-x_k)w(x-x_l)
 \end{equation}
we can omit the off-site terms since they are also typically two
orders of magnitude smaller than the on-site interaction matrix
elements. Note that $g_{1D}$ is the one-dimensional on-site
interaction strength, originating from an adjustment of the
scattering length $a_s$, due to the transversal
trapping~\cite{olshanii98}. As a central result of our studies we
therefore obtain a generalized Bose-Hubbard Hamiltonian:
\begin{eqnarray}\label{BHham2} \nonumber H&=& E_0 \hat{N}+E\hat{B}+\left( \hbar U_0 a^\dag a+V_{\textrm{cl}}\right)\left( J_0\hat{N}+ J\hat{B}\right) \\\nonumber&+&\hbar\eta_{\textrm{eff}} \left( a+a^\dag\right)
     \tilde{J}_{0}\sum_{k}(-1)^{k+1}\hat{n}_k-\hbar\Delta_c a^\dag a\\&-&i\hbar\eta\left( a-a^\dag\right)+\frac{U}{2}\sum_k\hat{n}_k\left(\hat{n}_k-1\right),
  \end{eqnarray}
where the nonlinear on-site interaction is characterized by
$U=g_{1D}\int\mathrm{d} x\left\vert w(x)\right\vert^4.$ In addition,
we introduced the number operator $\hat{N}=\sum_k\hat{n}_k=\sum_k
b_k^\dag b_k$ and the jump operator $\hat{B}=\sum_{k}\left(
b_{k+1}^\dag b_k+h.c.\right) $. Note that for strong classical
intracavity fields and no transverse pump we recover the standard
Bose-Hubbard Hamiltonian.

Finally, let us remark that we now can also rewrite the field Heisenberg Eq.~\eqref{eq:heisenberg} in the above terms, which gives:
\begin{multline}\label{eq:heisenberg2}
\dot{a}=\left\{i\left[ \De_c-U_0\left(
J_0\hat{N}+J\hat{B}\right)\right]-\kappa\right\}a+\eta\\-i\eta_{\textrm{eff}}\tilde{J}_0\sum_k(-1)^{k+1}\hat{n}_k.
\end{multline}
Here we clearly see that besides the number operator $\hat{N}$ for
the atoms also the coherence properties via the operator $\hat{B}$
and statistics via $\hat{n}_k$ play a decisive role in the field
dynamics. As this field acts back on the atomic motion, interesting
and complex coupled dynamics can be expected from this model, which
was partly already discussed in~\cite{maschler05,lewenstein06} and
will be elucidated more in the remainder of this work.
\section{Cavity pump}\label{sec:cav_pump}

Let us now turn to the conceptually simplest case and restrict the
pumping only to the cavity, where only a single mode is coherently
excited ({\it cavity pumping}). This mode will generate an optical
potential in addition to the prescribed external potential. For
large enough photon numbers the external potential can even be
omitted and the particles are trapped solely by the cavity field. As
essential ingredient in the dynamics, the identical coupling of all
atoms to this same field mode induces a long-range interaction
between the atoms independent of their positions. Setting
$\eta_{\textrm{eff}}=0$, the Hamiltonian~\eqref{BHham2} is reduces
to:
\begin{eqnarray}\label{BHhamCP} H&=& E_0 \hat{N}+E\hat{B}+\left( \hbar U_0 a^\dag a+V_{\textrm{cl}}\right)\left( J_0\hat{N}+ J\hat{B}\right) \nonumber\\&-&\hbar\Delta_c a^\dag a-i\hbar\eta\left( a-a^\dag\right)+\frac{U}{2}\hat{C}.
  \end{eqnarray}
Here we introduced $\hat{C}=\sum_k\hat{n}_k\left(\hat{n}_k-1\right)$
for the operator of the two-body on-site interaction. Still we see that
the corresponding Heisenberg equation for the cavity field:
\begin{equation}\label{eq:heisenbergCP}
\dot{a}=\left\{i\left[ \De_c-U_0\left(
J_0\hat{N}+J\hat{B}\right)\right]-\kappa\right\}a+\eta.
\end{equation}
depends on atom number and coherence. For very weak fields this
yields an atom statistics dependent cavity transmission spectrum,
which was studied in some detail in Ref.~\cite{mekhov2}. Here we go
one step further and study the dynamical back action of the field
onto atomic motion and field mediated atom-atom interaction, which
appear at higher photon number. As the model is still rather complex
we need some further approximations at this point in order to catch
some qualitative insight.
\subsection{Field-eliminated Hamiltonian}\label{sec:field_elim_Ham}

Although the influence of the cavity field on the atoms is equal on all particles, their common interaction
generates a dynamics much more complex than for a Bose-Hubbard model with prescribed external potential. This is more analogous to real solid state physics where the state of the electrons also acts back on the potentials. To exhibit the underlying physics, we
will now derive an approximate Hamiltonian, which solely depends on particle variables by adiabatically eliminating the field~\eqref{BHhamCP}. This should be valid when the damping rate $\kappa$ of the cavity generates a faster time scale than the external atomic
degrees of freedom. Actually as tunneling is mostly a very slow process (much slower than the recoil frequency), this will be almost always the case in practical experimental setups. To this end, we simply equate~\eqref{eq:heisenbergCP} to zero and obtain formally
$a=\eta/\{\kappa-i[\Delta_c-U_0(J_0\hat{N}+J\hat{B})]\}$. In the
following we constrain ourselves to the case of a fixed number of
atoms, i.e., $\hat{N}=N\mathbf{1}$. The very small tunneling matrix
element $J$ can be used as an expansion parameter, leading to:
\begin{equation}\label{eq:elimfield}
a\approx\frac{\eta}{\kappa-i\Delta_c^\prime}\left[\mathbf{1}-i\frac{U_0J}{\kappa-i\Delta_c^\prime}\hat{B}-\frac{(U_0J)^2}{(\kappa-i\Delta_c^\prime)^2}\hat{B}^2\right],
\end{equation}
where we introduced a shifted detuning
$\Delta_c^\prime=\Delta_c-U_0J_0N$.

In order to obtain an effective Hamiltonian, where the cavity
degrees of freedom are eliminated, we replace the field terms
in~\eqref{BHhamCP}, by the steady state
expressions~\eqref{eq:elimfield}, as well as in the Liouville
super operator~\eqref{eq:liouvillian}. Note, that this is more
appropriate than the naive approach of a replacement just in the Hamiltonian, as has been done in our former work~\cite{maschler05}. If
we consider terms up to order $\propto J^2$, the exchange in the
Hamiltonian yields:
\begin{multline}\label{eq:adhamproc2}
H_{\textrm{ad}}=(E+JV_{\textrm{cl}})\hat{B}+\frac{U}{2}\hat{C}\\+\frac{\hbar U_0 J \eta^2}{\kappa^2+{\Delta_c^\prime}^2}\left(\frac{{\Delta_c^\prime}^2-\kappa^2}{\kappa^2+{\Delta_c^\prime}^2}\hat{B}-\frac{3U_0J\Delta_c^\prime}{\kappa^2+{\Delta_c^\prime}^2}\hat{B}^2\right).
\end{multline}
Next, by applying the same procedure to the Liouville equation -
again up to terms~$\propto J^2$ -  we obtain an adiabatic Liouville
operator:
\begin{multline}\label{eq:adhamproc3}
\mathcal{L}_{\textrm{ad}}\varrho = -i\frac{2U_0J\kappa^2\eta^2}{\left(\kappa^2+{\Delta_c^\prime}^2\right)^2}\left[\hat{B}+\frac{2\Delta_c^\prime U_0J}{\kappa^2+{\Delta_c^\prime}^2}\hat{B}^2,\varrho\right]\\+
\frac{\kappa U_0^2J^2\eta^2}{\left(\kappa^2+{\Delta_c^\prime}^2\right)^2}\left(2\hat{B}\varrho\hat{B}-\hat{B}^2\varrho-\varrho\hat{B}^2\right).
\end{multline}
The Lindblad terms in the second line are real, corresponding to
dissipation, whereas the first, imaginary term corresponds to a
unitary time evolution and has therefore to be added to the adiabatic
Hamiltonian, i.e.,
\[ H_{\textrm{ad}}\rightarrow H_{\textrm{ad}} + \frac{2\hbar U_0J\kappa^2\eta^2}{\left(\kappa^2+{\Delta_c^\prime}^2\right)^2}\left(\hat{B}+\frac{2\Delta_c^\prime U_0J}{\kappa^2+{\Delta_c^\prime}^2}\hat{B}^2\right).\]
Altogether, we end up with a Hamiltonian, where the cavity field has been eliminated:
\begin{multline}\label{eq:effHam0}
H_{\textrm{ad}}=(E+JV_{\textrm{cl}})\hat{B}+\frac{U}{2}\hat{C}\\+\frac{\hbar U_0 J \eta^2}{\kappa^2+{\Delta_c^\prime}^2}\left(\hat{B}+\frac{U_0J \Delta_c^\prime }{\kappa^2+{\Delta_c^\prime}^2}\frac{\kappa^2-3{\Delta_c^\prime}^2}{\kappa^2+{\Delta_c^\prime}^2}\hat{B}^2\right).
\end{multline}
The loss rate of the cavity is described by the remaining dissipative part of~\eqref{eq:adhamproc3}:
\begin{equation}\label{eq:effLiou}
\mathcal{L}_{\textrm{ad}}\varrho = \frac{\kappa U_0^2J^2\eta^2}{\left(\kappa^2+{\Delta_c^\prime}^2\right)^2}\left(2\hat{B}\varrho\hat{B}-\hat{B}^2\varrho-\varrho\hat{B}^2\right).
\end{equation}

Note, that the above adiabatic elimination procedure is not completely unambiguous due to ordering freedom. Nevertheless it should give a qualitatively correct first insight. An alternative way of deriving an effective Hamiltonian, depending solely on particle observable is similar to~\eqref{eq:effHam2ndquant} and~\eqref{ad_ham}. This amounts to a replacement of the field
variables with~\eqref{eq:elimfield} in the Heisenberg equation for the external atomic degrees of freedom, which read as follows:
\begin{equation}\label{eq:particledyn}
\dot{b}_k = \frac{1}{i\hbar}\left[\left(E+JV_{cl}+\hbar U_0Ja^\dagger a\right)\left(b_{k-1}+b_{k+1}\right)-U\hat{n}_kb_k\right].
\end{equation}
A naive replacement of the field operator $a$ and its adjoint $a^\dagger$ by~\eqref{eq:elimfield} in the above expression leads to an equation for $\dot{b}_k$, which cannot be generated from an effective adiabatic Hamiltonian in the form $\dot{b}_k=-i/\hbar[b_k,H_{\textrm{ad}}]$.  Hence, before substituting the adiabatic field operators, we have to symmetrize the expression
containing the field term in~\eqref{eq:particledyn} in the form
\begin{multline}\label{eq:particledyn2}
\dot{b}_k = -\frac{i}{\hbar}\left[\left(E+JV_{cl}\right)\left(b_{k-1}+b_{k+1}\right)-U\hat{n}_kb_k\right]\\
-\frac{i\hbar U_0J}{2}\left[a^\dagger a\left(b_{k-1}+b_{k+1}\right)+\left(b_{k-1}+b_{k+1}\right)a^\dagger a\right].
\end{multline}
This form enables us to describe the dynamics of $b_k$ by a Heisenberg equation with an effective Hamiltonian, which up to second order in $J$ reads:
\begin{multline}\label{effHam}
H_{\textrm{ad}}=(E+JV_{\textrm{cl}})\hat{B}+\frac{U}{2}\hat{C}\\+\frac{\hbar U_0 J \eta^2}{\kappa^2+{\Delta_c^\prime}^2}\left(\hat{B}+\frac{U_0J\Delta_c^\prime}{\kappa^2+{\Delta_c^\prime}^2}\hat{B}^2\right).
\end{multline}
The terms in the second line stem from the field terms in~\eqref{eq:particledyn2}.
Although this Hamiltonian looks a bit different from the first version derived before~\eqref{eq:effHam0}, their properties are - within their regime of
validity - in very good agreement as long as hopping is slow compared to damping.

\begin{figure}[htp]
    \centering
     \includegraphics[width=0.8\hsize]{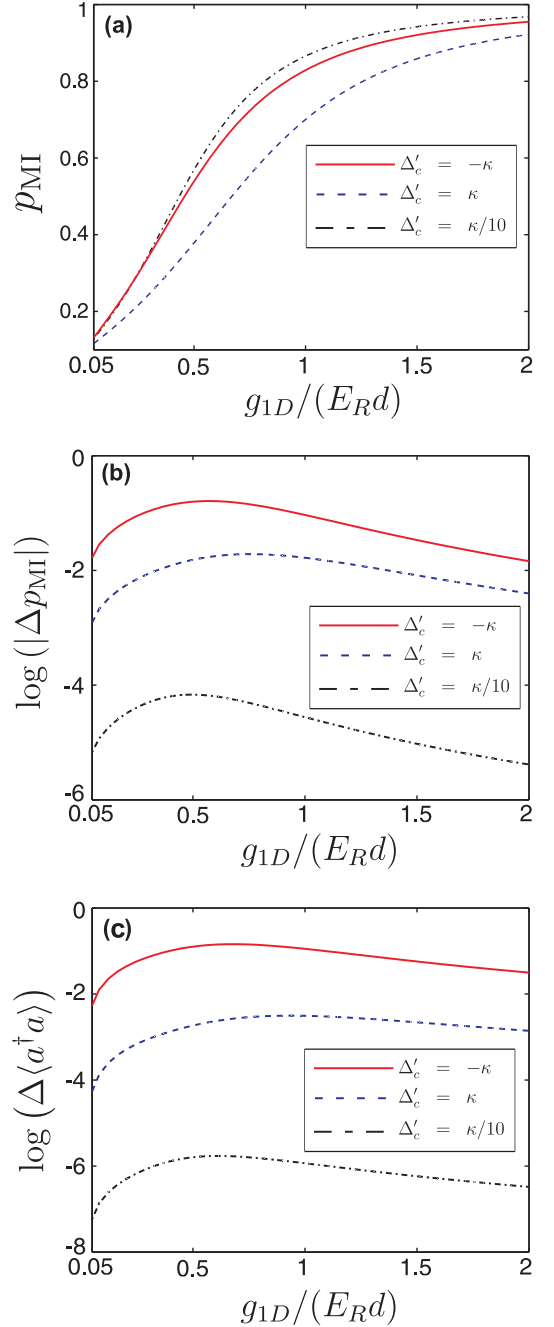}
     \caption{\label{fig:ham_diff}(color online) (a) Contribution of the Mott-insulator state to the ground state $p_{\textrm{MI}}=\vert \langle \psi\vert\textrm{MI}\rangle\vert^2$ of~\eqref{effHam} as function of the 1D on-site interaction strength in units of $E_Rd$ ($d$ is the lattice constant) (b) Logarithmic difference of  $p_{\textrm{MI}}$, calculated with the groundstate of~\eqref{effHam} and~\eqref{eq:effHam0}.
     (c) Logarithmic difference of the adiabatically eliminated photon number $\langle \psi\vert a^\dagger a\vert\psi\rangle$ with $a$ from~\eqref{eq:elimfield}
     for the two different ground states. The parameters are  $\kappa=1/\sqrt{2}\omega_R,\eta=2.35\omega_R$ and $\Delta_c^\prime=-\kappa$ (red, solid line), $\kappa=4\omega_R,\eta=12.5\omega_R$ and
     $\Delta_c^\prime=\kappa$ (blue, dashed line) and $\kappa=4\omega_R,\eta=10\omega_R$ and
     $\Delta_c^\prime=-\kappa/10$ (black, dashed-dotted line).
     In any of the curves, we set $V_{\textrm{cl}}=0$ and
     $U_0=-\omega_R$. Here $\omega_R$ is the frequency corresponding to the recoil energy, i.e., $E_R=\hbar^2 k^2/(2m)=\hbar\omega_R$.}
    \nobreak\medskip
  \end{figure}

To exhibit the physical content of this Hamiltonian one can look at
its eigenstates. As first step we calculate the Mott insulator state
[see Eq.~\eqref{eq:Mottstate}] fraction of the lowest energy state
$\vert\psi\rangle$ of these two Hamiltonians, i.e.,
$p_{\textrm{MI}}=\vert \langle \psi\vert\textrm{MI}\rangle\vert^2$
(see also Fig.~\ref{fig:phase_trans}), as a function of the on-site
interaction energy for different values of $\Delta_c^\prime$. This
will indicate changes of position and behavior of the Mott insulator
superfluid transition (see Fig.~\ref{fig:phase_trans}). To compare
the two approximate Hamiltonians in Fig.~\ref{fig:ham_diff}, we plot
the difference of the Mott insulator fraction of the ground state
of~\eqref{eq:effHam0} and~\eqref{effHam}, as well as the difference
of the steady state photon number. Obviously the two Hamiltonians,
converge in the limit of large cavity decay $\kappa$. This can also
be seen in Fig.~\ref{fig:ham_diff}, where the dashed-dotted line
depicts the case of a smaller $\Delta_c^\prime$ (which is equivalent
to an enlarged $\kappa$), showing a strongly enhanced coincidence.

\subsection{Field-eliminated density operator}

Let us now use a further and somehow more systematic alternative approach to eliminate the cavity field dynamics
from the system evolution directly from the Liouville equation by following a method proposed by Wiseman and
Milburn~\cite{wiseman93}, which is valid for large $\kappa$ and low photon numbers. In this case we have
\begin{equation}\label{bad_cav_approx}
\left\vert\frac{\langle H_{at}\rangle}{\kappa}\right\vert\sim\left\vert\frac{\hbar U_0\langle a^\dagger a\rangle}{\kappa}\right\vert=\varepsilon\ll1,
\end{equation}
where $H_{at}$ is the atomic part of~\eqref{BHhamCP}, i.e.,
$H_{at}=\left(E+V_{cl}J\right)\hat{B}+U\hat{C}/2$. Again the total atom number $\hat{N}$ is supposed
to be constant. This allows to expand the density operator in powers of $\varepsilon$, corresponding to states with increasing photon number:
\begin{eqnarray}
\varrho &=& \varrho_0\otimes \vert 0\rangle_a\langle0\vert+\left(\varrho_1\otimes\vert1\rangle_a\langle0\vert + h.c.\right)\\
&&+\varrho_2\otimes\vert 1\rangle_a\langle 1\vert +\left(\varrho_2^\prime\otimes\vert 2\rangle_a\langle 0\vert + h.c.\right) + O\left(
\varepsilon^3\right).\nonumber
\end{eqnarray}
Here $\varrho_i$ are density operators for the particle variables,
corresponding to the order $i$ of magnitude in the expansion
parameter $\varepsilon$. We substitute this expression into the
Liouville equation~\eqref{eq:liouville} with the Hamiltonian
from~\eqref{BHhamCP}, which leads to the following set of equations:
\begin{subequations}
\begin{eqnarray}
\label{eq1}\dot{\varrho}_0 &=& \frac{1}{i\hbar}\left[H_{at},\varrho_0\right] -\eta\left(\varrho_1+\varrho_1^\dagger\right) + 2\kappa\varrho_2 \\
\label{eq2}\dot{\varrho}_1 &=&
\frac{1}{i\hbar}\left[H_{at},\varrho_1\right]-\eta\left(\sqrt{2}\varrho_2^\prime+\varrho_2-\varrho_0\right)-\kappa\varrho_1
\\&
+& i\left[\Delta_c-U_0\left(J_0N+J\hat{B}\right)\right]\varrho_1+\kappa O\left(\varepsilon^4\right)\nonumber\\
\label{eq3}\dot{\varrho}_2 &=& \frac{1}{i\hbar}\left[H_{at},\varrho_2\right]+\eta\left(\varrho_1+\varrho_1^\dagger\right) - 2\kappa\varrho_2 \\&-&
iU_0\left[J_0N+J\hat{B},\varrho_2\right]+\kappa O\left(\varepsilon^4\right)\nonumber\\
\label{eq4}\dot{\varrho}_2^\prime &=& \frac{1}{i\hbar}\left[H_{at},\varrho_2^\prime\right]+\sqrt{2}\eta\varrho_1 - 2\kappa\varrho_2^\prime\\
&+& 2i\left[\Delta_c-U_0\left(J_0N+J\hat{B}\right)\right]\varrho_2^\prime +\kappa O\left(\varepsilon^4\right)\nonumber,
\end{eqnarray}
\end{subequations}
Now we adiabatically eliminate the off-diagonal elements $\varrho_1$
and $\varrho_2^\prime$. Setting their derivations in~\eqref{eq2}
and~\eqref{eq4} to zero and neglecting terms with respect to the
assumption~\eqref{bad_cav_approx}, we obtain:
\begin{equation}\label{eq4b}
  \varrho_2^\prime=\frac{\eta}{\sqrt{2}A}\varrho_1 + O\left(\varepsilon ^3\right).
\end{equation}
This is consistent with the assumption $\varrho_2^\prime\sim
O\left(\varepsilon^2\right)$. Here we defined
$A=\kappa-i\Delta_c^\prime +iU_0J\hat{B}$.
Putting~\eqref{eq4b} into~\eqref{eq2} and neglecting the terms
consistent with the order of the expansion, such that $\varrho_1\sim
O\left(\varepsilon\right)$, it follows that:
\begin{equation}
\varrho_1 = \frac{\eta}{A+\eta^2/A}\left(\varrho_0-\varrho_2\right) + O\left(\varepsilon ^4\right).
\end{equation}
We simplify this expression, $\varrho_1\approx
\eta{A}^{-1}(\varrho_0-\varrho_2)$, which is consistent with the
above expansion and substitute it into~\eqref{eq1} and~\eqref{eq3}:
\begin{subequations}
\begin{eqnarray}
\label{eq1b}\dot{\varrho}_0 &=& \frac{1}{i\hbar}\left[H_{at},\varrho_0\right] + 2\kappa\varrho_2\nonumber \\&-&\eta^2\left[A^{-1}
\left(\varrho_0-\varrho_2\right)+\left(\varrho_0-\varrho_2\right){A^\dagger}^{-1}\right]\\
\label{eq3b}\dot{\varrho}_2 &=& \frac{1}{i\hbar}\left[H_{at},\varrho_2\right] - iU_0\left[J_0N+J\hat{B},\varrho_2\right] - 2
\kappa\varrho_2\nonumber \\&+& \eta^2\left[A^{-1}\left(\varrho_0-\varrho_2\right)+\left(\varrho_0-\varrho_2\right){A^\dagger}^{-1}\right].
\end{eqnarray}
\end{subequations}
In order to formulate a master equation for the particle variables
we have to use the reduced density operator, where we trace over the
field variables, i.e., $\varrho_{at} = \textrm{tr}(\varrho) =
\varrho_0+\varrho_2 + O\left(\varepsilon^4\right)$.
With~\eqref{eq1b} and~\eqref{eq3b} we see that:
\begin{equation}\label{eq:mastereq}
\dot{\varrho}_{at} = \frac{1}{i\hbar}\left[H_{at},\varrho_{at}\right] - iU_0\left[J_0N+J\hat{B},\varrho_2\right].
\end{equation}
As a further approximation, which is also consistent with the
expansion order of the assumption~\eqref{bad_cav_approx}, we
set~\eqref{eq3b} to zero and neglect $\left[H_{at},\varrho_2\right]$
and all other terms smaller than $O\left(\varepsilon^3\right)$. Then
we can express $\varrho_2$ through $\varrho_0$:
\begin{equation}
\varrho_2=\frac{\eta^2}{2\kappa}\left[A^{-1}\varrho_0+\varrho_0{A^\dagger}^{-1}\right].
\end{equation}
Within this order of magnitude of $\varepsilon$ we can replace
$\varrho_0$ with $\varrho_{at}$, leading us finally to the following
master equation for the reduced density operator of the particle
variables:
\begin{eqnarray}\label{eq:elim_liouville}
\dot{\varrho}_{at}&=&\frac{1}{i\hbar}\left[H_{at},\varrho_{at}\right]\\&&\nonumber - i\frac{U_0\eta^2}{2\kappa}\left[J_0N+J\hat{B},
\left(A^{-1}\varrho_{at}+\varrho_{at}{A^\dagger}^{-1}\right)\right].
\end{eqnarray}
Note that this model also contains a damping part, since the
operator $A$ is not hermitian. Let us investigate this damping, by
expanding the inverse of $A$ up to first order in $J$, which
is consistent with the order of magnitude
in~\eqref{eq:elim_liouville}. Hence we replace $A^{-1}$ and its
adjoint in this equation by
\begin{equation}\label{approxA}
A^{-1}\approx \frac{1}{\kappa-i\Delta_c^\prime}\left(1-i\frac{U_0J}{\kappa-i\Delta_c^\prime}\hat{B}\right)
\end{equation}
and its adjoint. Since we are restricted on a subspace of constant
atom number, the Liouville equation reads as
follows:
\begin{eqnarray}
\label{eq:elim_liouville2}
\dot{\varrho}_{at}&=&\frac{1}{i\hbar}\left[H_{at} +\frac{\hbar U_0\eta^2}{\kappa^2+{\Delta_c^\prime}^2}\left(
J\hat{B}+\frac{U_0\Delta_c^\prime J^2}{\kappa^2+{\Delta_c^\prime}^2}\hat{B}^2\right),\varrho_{at}\right]\nonumber \\
&&-\frac{(JU_0\eta)^2}{2\kappa}\frac{\kappa^2-{\Delta_c^\prime}^2}{\left(\kappa^2+{\Delta_c^\prime}^2\right)^2}
\left[\hat{B},\left[\hat{B},\varrho_{at}\right]\right].
\end{eqnarray}
Obviously, the non-dissipative part of this equation agrees perfect
with our adiabatically eliminated Hamiltonian~\eqref{effHam} and the
structure of the dissipative part is of the same Lindblad form
as~\eqref{eq:effLiou}. Note that an expansion of $A^{-1}$ to higher
order in $J$ would also provide us the correct next-order term
of~\eqref{effHam} plus an extra term in the Liouville-equation,
which does not correspond to unitary time evolution, as described by
a Hamiltonian. This confirms the usefulness of the naive elimination
method, also used in Ref.~\cite{maschler05}.

\subsection{Quantum phase transitions in an optical lattice}\label{sec:QPT_CP}

\begin{figure}[htp]
    \begin{center}
     \scalebox{0.35}[0.35]{\includegraphics{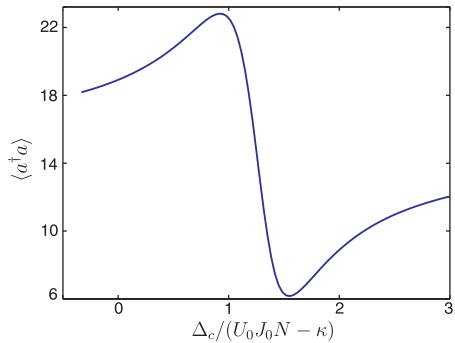}}
     \caption{\label{fig:deltaC_var}(color online). Self-consistent photon number in the case of four particles in four wells
     without on-site interaction. Parameters are $U_0=-\omega_R$ and $\kappa=\omega_R$.}
    \nobreak\medskip
   \end{center}
  \end{figure}

In section~\ref{sec:field_elim_Ham} we derived two approximate
Hamiltonians~\eqref{eq:effHam0} and~\eqref{effHam} describing our
system of cold atoms in an optical lattice. To a large extend they
still implement the well known BH model, but with parameters
controllable via cavity detuning and some additional nonlocal
interaction terms. Let us now investigate their properties in some
more detail. One of the key features of optical cavities is the
feedback mechanism between atoms and cavity field. Hence,
computations are a subtle issue, since the matrix elements in the BH
Hamiltonian depend on the field amplitude, which itself depends on
the atomic positions. In principle a rigorous treatment would
consist of calculating the matrix elements~\eqref{mat_els} for every
photon Fock state and treating the parameters of the BH model as
operators. To avoid the full complexity of such an approach we will
first assume only a weak dependence of the Wannier functions on the
mean cavity photon number $\langle a^\dagger a\rangle$, which allows
us to proceed analytically. For any set of operating parameters we
then calculate the matrix elements in a self-consistent way
replacing the photon number operator by its average in the iteration
process. Explicitly this is implemented by starting from some
initial guess $J_0^{(0)},\,E_0^{(0)},\,J^{(0)},\,E^{(0)}$ in the
Hamiltonian~\eqref{effHam}, from which we calculate the ground state
$\vert \psi^{(0)}\rangle$. By use of this state we obtain an initial
mean photon number $\langle\psi^{(0)}\vert a^\dag a\vert
\psi^{(0)}\rangle$, with the steady-state field
operator~\eqref{eq:elimfield}. Now we can calculate the matrix
elements $J_0^{(1)},\,E_0^{(1)},\,J^{(1)},\,E^{(1)}$ again leading
to a new ground state $\vert \psi^{(1)}\rangle$ and a new mean
photon number $\langle\psi^{(1)}\vert a^\dag a\vert
\psi^{(1)}\rangle$. Proceeding iteratively, in most cases the
fixpoint is reached already after very few iterations and the system
properties are then calculated with this self-consistent matrix
elements. The convergence speed decreases near the resonance for the
cavity photon number (cf. Fig.~\ref{fig:deltaC_var}), which occurs
for $\Delta_c=U_0J_0N-\kappa$, especially for large $U_0$.
Introducing some damping in the iteration procedure easily resolves
this issue, though.
As we mentioned already before, we restrict the model on a subspace
$\mathcal{H}_N$ of a fixed total particle number $N$ in an optical
lattice of $M$ sites. A basis of $\mathcal{H}_N$ consists of the
states $\vert N,0,0,\ldots,0\rangle,\vert
N-1,1,0,\ldots,0\rangle,\ldots,\vert 0,0,\ldots,0,N\rangle.$ Since
we are interested in the quantum phase transition between the Mott
insulator (MI) and the superfluid (SF) state occurring during the
variation of certain external parameters, we investigate the
contributions of these specific states to the ground state of the
atomic system. The Mott insulator state is a product of Fock states
with uniform density distribution, i.e.,
\begin{equation}\label{eq:Mottstate}
\vert \mathrm{MI}\rangle=\vert n,n,\ldots,n\rangle,
\end{equation}
 with $n=N/M$. In contrast, in a SF state each atom is delocalized over all sites. It is given by a superposition of Fock states,
 namely of all possible distributions of the atoms in the lattice sites, i.e.,
\begin{equation}
\vert \mathrm{SF}\rangle = \sum_{k_1,k_2,\ldots,k_M}\frac{N!}{\sqrt{M^N}\sqrt{k_1!k_2!\cdots k_M!}}\vert k_1,k_2,\ldots,k_M\rangle,
\end{equation}
 with $\sum_{i=1}^M k_i=N$. Although the density in the superfluid state is also uniform $\langle \hat{n}_i\rangle_{\mathrm{SF}}=N/M$
and therefore equal to the Mott insulator state, its properties are fundamentally different. This manifests especially in the spectra and angle dependence of scattered light, providing for new, non-destructive probing schemes for the atomic phases~\cite{mekhov1,mekhov2}.

\begin{figure}[htp]
    \begin{center}
      \scalebox{0.75}[0.75]{\includegraphics{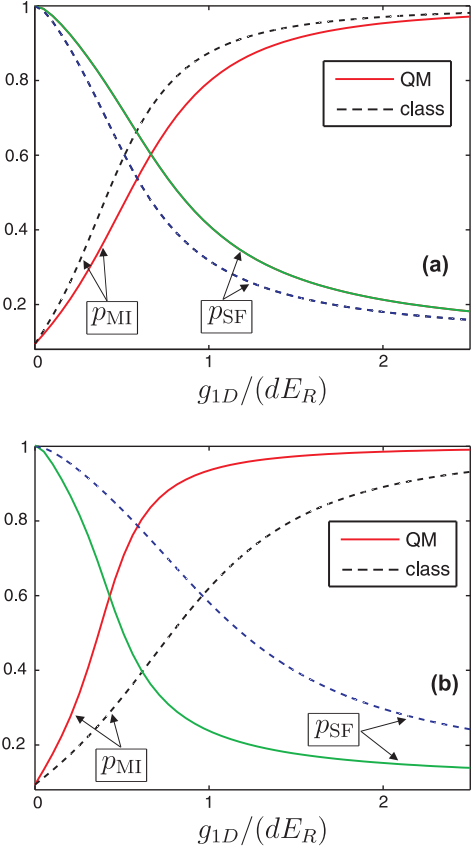}}
     \caption{\label{fig:phase_trans}(color online) Cavity influence of the Mott insulator to superfluid transition by means of a comparison of the
     occupation probabilities $p_{\textrm{MI}}$ and $p_{\textrm{SF}}$ for a purely quantum field, i.e., $V_{cl}=0$, and a purely classical field, i.e., $\eta=0$, as a function of the dimensionless one-dimensional on-site interaction strength $g_{1D}/(dE_R)$. We choose $\eta$ such that both potentials are of equivalent depth, $V=5.5E_R$, for zero on-site interaction ($g_{\textrm{1D}}=0$). The quantum and classical case is
     depicted with solid and dashed lines, respectively. In (a) we set $(U_0,\kappa,\eta)=(-1,1/\sqrt{2},\sqrt{5.5})\omega_R$ and
     $\Delta_c-U_0J_0N=\kappa$. (b) The same as (a) but with $\Delta_c-U_0J_0N=-\kappa$.}
    \nobreak\medskip
   \end{center}
  \end{figure}

Let us now investigate the influence of the cavity on position and
shape of the well-known ``classical''
MI-SF-transition~\cite{Jaksch98,Zwerger03,Fisher89}. To do so, we
compare the two cases of a pure quantum field, i.e., $V_{cl}=0$
in~\eqref{effHam}, and a classical field ($\eta=0$) provided by
$V_{cl}$ for generating the optical potential. We choose $\eta$ in
such a way, that at zero on-site interaction, $g_{\textrm{1D}}=0$,
both potentials are equally deep. As depicted in
Fig.~\ref{fig:phase_trans}, the influence of the cavity strongly
depends on the detuning $\Delta_c$. Two contributions arise from the
quantum nature of the potential. On the one hand the potential depth
and therefore the matrix elements depend on the atomic state. For a
classical potential this is clearly not the case. On the other hand
the cavity mediates long-range interactions via the field, which
corresponds to the $\hat{B}^2$-term in~\eqref{effHam}. If a
potential depth near the phase transition point for the quantum case
is associated with some certain average photon number $\bar{n}$,
then $\bar{n}\pm 1$ are associated with different atomic phases.
This means that the ground state of the quantized cavity field
contains contributions of different atomic states, each of them
correlated with the corresponding photon number. In this sense
photon number fluctuations drive particle fluctuations. Depending on
parameters the former or the latter effect contributes more. In
Fig.~\ref{fig:phase_trans} this is shown for four atoms in four
wells, where we calculated the occupation probability for the Mott
insulator
$p_{\textrm{MI}}=\vert\langle\psi\vert\textrm{MI}\rangle\vert^2$ and
the superfluid state
$p_{\textrm{MI}}=\vert\langle\psi\vert\textrm{SF}\rangle\vert^2$ for
the ground state $\vert\psi\rangle$ of~\eqref{effHam} as a function
of the dimensionless one-dimensional on-site interaction strength
$g_{1D}/(dE_R)$ for a purely classical and a purely quantum case.
For $\Delta_c-U_0J_0N=\kappa$, photon number fluctuations enhance
particle fluctuations, shifting the superfluid to Mott insulator
transition to higher values of the on-site interaction
[Fig.~\ref{fig:phase_trans}(a)]. However, if we choose
$\Delta_c-U_0J_0N=-\kappa$, the influence of the atomic state on the
potential depth exceeds the cavity-mediated long-range interactions,
strongly shifting the transition to lower values of $g_{1D}$
[Fig.~\ref{fig:phase_trans}(b)]. Note, that for this behavior, the
cavity loss rate must be - although within the bad cavity limit -
small enough. For larger $\kappa$ the quantum effects disappear and
the ground states for classical and quantum potential coincide.

\begin{figure}[htp]
    \begin{center}
      \scalebox{0.35}[0.35]{\includegraphics{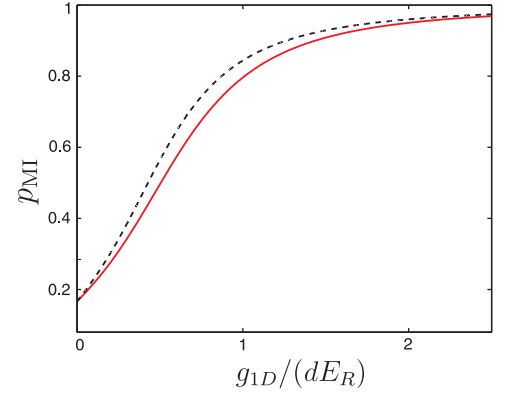}}
     \caption{\label{fig:longrange}(color online) Influence of the long-range interaction on the Mott insulator to superfluid phase transition,
     mediated via the $\hat{B}^2$ term in~\eqref{effHam}. The solid line shows the probability for the Mott insulator state as a function of
     dimensionless one-dimensional on-site interaction strength $g_{1D}/(dE_R)$ for a a purely quantum field, i.e., $V_{cl}=0$.
     The dashed line corresponds to the probability for the same Hamiltonian, neglecting the $\hat{B}^2$ term.
     The parameters are the same as in Fig.~\ref{fig:phase_trans}(b).}
    \nobreak\medskip
   \end{center}
  \end{figure}

To correctly address the long-range interactions, corresponding to
the $\hat{B}^2$ term in~\eqref{effHam}, we calculate the
contribution of the Mott insulator state to the ground state of this
adiabatic Hamiltonian including and omitting the $\hat{B}^2$ part,
respectively. Although, in the situation of
Fig.~\ref{fig:phase_trans}(b) the net effect enhances the phase
transition, the cavity mediates long-range coherence via
$\hat{B}^2$, which can be seen by enlarged particle number
fluctuations as shown in Fig.~\ref{fig:longrange}. Although the
effect is not too strong as it depends on $J^2$ is has infinite
range and will get more important for large particle numbers.

\begin{figure}[htp]
    \begin{center}
     \scalebox{0.35}[0.35]{\includegraphics{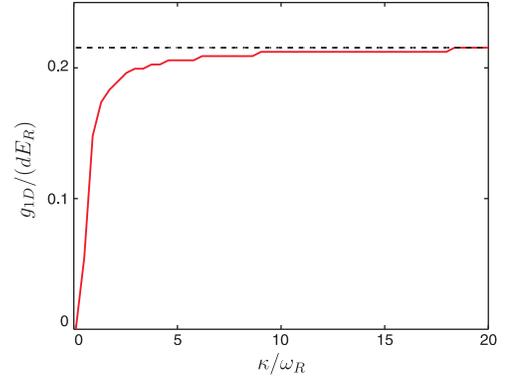}}
     \caption{\label{fig:qm2cl}(color online) Value of the on-site interaction $g_{1D}$, where the contributions of the Mott state and the superfluid state to the ground state of~\eqref{effHam} are equal, i.e., $\vert \langle \psi\vert\textrm{MI}\rangle\vert = \vert \langle \psi\vert\textrm{SF}\rangle\vert$, as a function of $\kappa$ (solid line) for a system of four atoms in four wells. Simultaneously we increase $\eta$, such that  $U_0\eta^2/\kappa^2=-6E_R$ is fixed. Obviously, the corresponding value at the same intersection point of a purely classical Bose-Hubbard Hamiltonian with
     $V_{cl}=-6E_R$ is constant (dashed line). Parameters are $U_0=-\omega_R,\,\kappa = 4\omega_R$ and $\Delta_C -U_0N=-\kappa$.}
    \nobreak\medskip
   \end{center}
  \end{figure}

Finally, we exhibit the transition from a cavity field with quantum
properties towards a classical optical lattice. This relies on the
assumption that a very bad cavity should be almost like no cavity
and increasing  $\kappa$, but keeping the potential depth constant,
approaches the classical limit. Hence, the effects of the quantum
nature and feedback of lattice potential should disappear and the
ground states for classical and quantum potential coincide. The
adiabatic eliminated Hamiltonian then has to approach the classical
Bose-Hubbard Hamiltonian. This is shown in Fig.~\ref{fig:qm2cl} for
a system of four atoms in four wells, where we simultaneously
increase $\kappa$ and $\eta$, keeping $U_0\eta^2/\kappa^2 = -6E_R$
fixed. For every $\kappa$ we calculated the value of the on-site
interaction $g_{\textrm{1D}}$, where the contributions of the Mott
state and the superfluid state to the ground state of~\eqref{effHam}
are equal, i.e., $\vert \langle \psi\vert\textrm{MI}\rangle\vert =
\vert \langle \psi\vert\textrm{SF}\rangle\vert$. This is compared
with the corresponding value of the interaction strength at the same
intersection point of a purely classical Bose-Hubbard model with a
potential depth of $V_{cl}=-6E_R$. We see that the transition occurs
already at a cavity linewidth of only an order of magnitude larger
than the recoil frequency, where the deviation is small already.
Thus one needs quite good resonators to see the quantum shift in the
phase transition.

\subsection{Comparison with the full dynamics of the master equation}

Using the approximate adiabatic model with eliminated field we have
found important changes in the physics so far. Even stronger effects
are to be expected in the limit of less and less cavity damping and
stronger atom field coupling. Let us now investigate some first
signs of this and test the range of validity of the above model in
this limit. To do so we have to resort to numerics and compare
solutions of the full master equation~\eqref{eq:liouville} with the
ground states of the adiabatically eliminated
Hamiltonian~\eqref{effHam}. Obviously solving the full master
equation is a numerically demanding task. Nevertheless, by
constraining to few atoms in few wells we are able to solve the
equations and reveal the essential physical mechanisms. The limit of
the band model description is of course reached for atoms coupled
strongly to a cavity field with only very few photons and no
additional classical potential $V_{cl}$ present. Here very strong
changes in the tunneling amplitudes occur whenever a photon leaks
out of the cavity and reduces the momentary potential depth. This
leads to strongly enhanced particle hopping. For instance, one can
think of the situations ``one photon present'' and ``no photon
present'', where the atoms can freely move within the cavity in the
absence of an external trap. On the other hand one extra photon can
almost block hopping. Note that in this case the ground state atomic
configuration can be close to superfluid for a low photon number and
close to an insulator state for a higher photon number. As our
matrix elements depend only on the mean photon number $\langle
a^\dagger a\rangle$, these differences cannot be taken into account
in an adiabatic model.

\begin{figure}[htp]
    \begin{center}
\includegraphics[width=0.8\hsize]{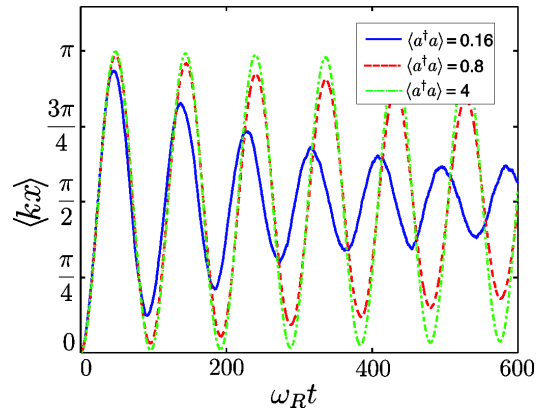}
     \caption{\label{fig:x_av_ph_nr_var}(color online) Mean position $\langle kx(t)\rangle$ of a single atom in two wells.
     We adjusted $U_0,\eta,\Delta_c$ in such a way, that the mean number of cavity photons increases, but the lattice depth stays nearly
     constant: $V=\hbar U_0\langle a^\dagger a\rangle=-8E_R$. Starting with $(U_0,\eta,\kappa)=(-50,10,25)$ (in units of $\omega_R$) and
     $\Delta_c = J_0U_0$, followed by a successive reduction of  $U_0$ by a factor of 5, together with an increase of $\eta$ by a factor of
     $\sqrt{5}$ and a proper adjustment of $\Delta_c$, this leads to mean photon numbers of 0.16 (solid line), 0.8 (dashed line), and 4
     (dashed-dotted line). Initially, the atom is in the right well.}
    \nobreak\medskip
   \end{center}
  \end{figure}

We can explicitly show this behavior by reducing the coupling
strength $U_0$, but keeping the average potential depth fixed (equal
matrix elements), by means of a higher average cavity photon number,
which leads to strongly reduced photon number fluctuations. The most
simple situation to discuss this issue is one atom loaded in a
lattice consisting of only two wells. Here, $\vert l\rangle\,(\vert
r\rangle)$ means the left (right) of the two wells, with a potential
minimum at $x=0 \,(x=\pi$). The hopping operator $\hat{B}$ then
describes tunneling from the left well to the right well and vice
versa. In Fig.~\ref{fig:x_av_ph_nr_var} we show this tunneling
behavior by plotting the mean position of the single atom $\langle
kx(t)\rangle$. The atomic ground state of this system is the
symmetric state $\vert\psi_0\rangle=\left(\vert l\rangle+\vert
r\rangle\right)/\sqrt2$ having a mean position of $\langle
kx\rangle_{\psi_0}=\pi/2$.  Decreasing $U_0$, increasing $\eta$  and
adjusting $\Delta_c$, yields different mean photon numbers $\langle
a^\dagger a\rangle$, but equal average lattice potential depth
$V=U_0\langle a^\dagger a\rangle$. (We do not consider an additional
classical potential here.) If only few photons are present, we
observe large fluctuations of the field and the system damps fast to
the ground state. As the photon number increases, the potential
approximates a classical potential as expected, where there is no
dephasing. The (nearly) equal oscillation frequencies show that the
matrix elements coincide for the different photon numbers. This is
an interesting feature corresponding to the quantum nature of the
potential. In contrast to the Bose-Hubbard model for a classical
optical lattice, lattice depth and interaction strength are not the
only important system parameters. Quantum fluctuations of the
potential are an additional source of atomic fluctuations, playing
an essential role in the evolution of the system. Obviously, if only
an external potential is present and the atom is no longer coupled
to the cavity field ($U_0=0$), the system is undamped, due to the
lack of the only dissipation channel present, cavity loss. In this
case the Hamiltonian~\eqref{BHhamCP} reduces to
$H=(E+JV_{cl})\hat{B}+U/2\hat{C}$, and the atom, initially not in
the symmetric state, oscillates between the left and right well.
Note that a more rigorous treatment of operator-valued matrix
elements - as described in the previous section - would be capable
of describing this behavior correctly. Alternatively for few atoms
Monte-Carlo wave function simulations of the full Hamiltonian could
be performed, allowing for processes, where the particle leaves the
lowest band~\cite{Vukics07}.

\begin{figure}[htp]
   \begin{center}
   \includegraphics[width=0.8\hsize]{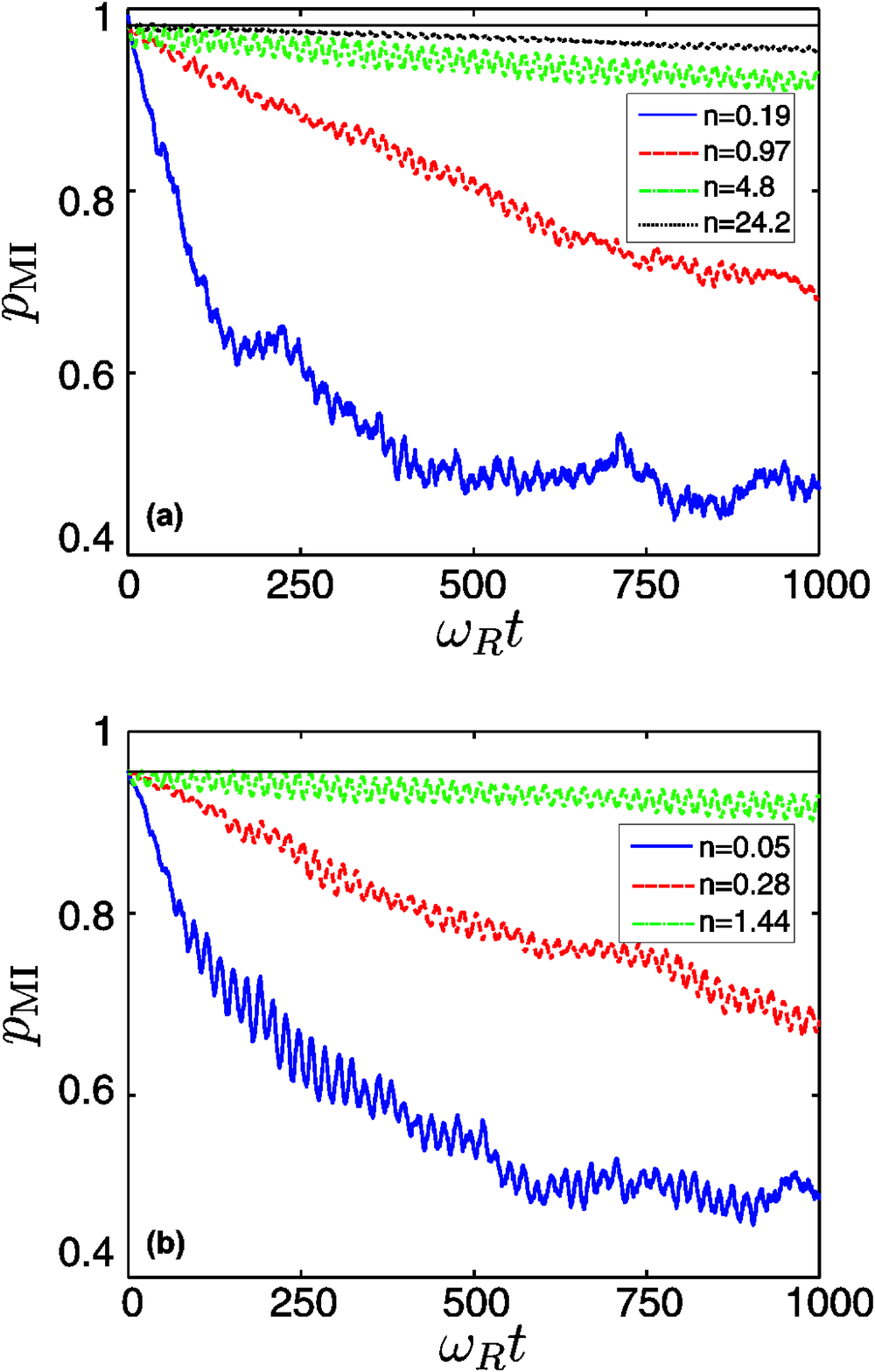}
  \caption{\label{fig:2atoms_phnr_var}(color online) Probability of the Mott insulator state $\vert\psi_{\textrm{MI}}(t)\vert^2$
  for two atoms in two wells. Parameters and procedure as in Fig.~\ref{fig:x_av_ph_nr_var}, but due the second atom the photon numbers are increased.
 The on-site interaction is $U=0.32E_R$. (a) $V_{cl}=0$. The curves correspond to a mean photon
number of 0.19 (solid line), 0.97 (dashed line), 4.8 (dashed-dotted
line) and 24.2 (dotted line). (b) $V_{cl}=-5E_R$ and corresponding photon numbers of
0.05 (solid line), 0.28 (dashed line), 1.44 (dashed-dotted line).}
    \nobreak\medskip
   \end{center}
  \end{figure}

Obviously, this enhancement of atom fluctuations for low photon
numbers also affects the dynamics of several atoms. We demonstrate
this for the case of two atoms in two wells. We assume strong
coupling with few cavity photons and a strong on-site interaction,
which - in principle - inhibits tunneling and drives the system
deeply into the Mott insulator regime. However, starting from a state
slightly perturbed from the ground state of the adiabatically
eliminated Hamiltonian~\eqref{effHam}, the system does not evolve
towards this Mott-like ground state but towards some other,
drastically different state. Increasing the photon number, while
keeping the lattice depth constant, reduces the atom fluctuations
and keeps the system near its adiabatic ground state. This is shown
in Fig.~\ref{fig:2atoms_phnr_var}(a), where the probability for the
system being in the Mott insulator regime
$p_{\textrm{MI}}=\vert\psi_{\textrm{MI}}(t)\vert^2$ is plotted.
Again we observe that, the larger the intracavity photon number is,
the more the potential approaches a purely classical one and the
more significant the ground state probabilities of~\eqref{effHam}
are. Hence we see that including the photon number fluctuations
strongly suppresses the Mott insulator state by allowing the particles to hop
during photon number fluctuations. This is also a strong restriction for the use of
our adiabatic model Hamiltonian, where only average photon numbers enter the
model parameters.


Clearly, some added external classical potential diminishes this problem
as it can ensure the existence of a bound state, independent of the number of cavity
photons, giving an upper limit to the hopping rate. This is demonstrated
in~Fig.~\ref{fig:2atoms_phnr_var}(b), where a classical potential of
$V_{cl}=-5E_R$ is added. Here for $\langle a^\dagger a\rangle=1.44$
the deviations from the adiabatic ground state are of the same order
as for  $V_{cl}=0$ for $\langle a^\dagger a\rangle=4.8$
[Fig.~\ref{fig:2atoms_phnr_var}(a)]. Nevertheless, for not too leaky
cavities ($\kappa$ is in an intermediate regime), the regime of
validity of the adiabatically eliminated Hamiltonian~\eqref{effHam}
is limited to case where either a large purely classical
potential or a large photon number is given.

\begin{figure}[htp]
    \begin{center}
\includegraphics[width=1\hsize]{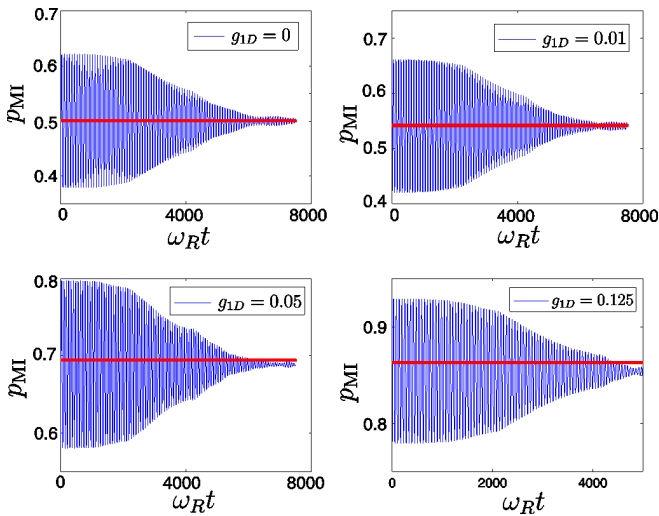}
  \caption{\label{fig:onsitevar}(color online) Probability of the Mott insulator state $\vert\psi_{\textrm{MI}}(t)\vert^2$ for two atoms in
  two wells for different on-site interaction. In (a) there is no interaction, i.e., $U=0$, in (b) $U=0.0065E_R$, in (c) $U=0.0324E_R$ and in
  (d) $U=0.081E_R$. Other parameters are $(U_0,\kappa,\eta,\Delta_c)=(-50,25,1,0)$ (in units of $\omega_R$), the classical potential is $V_{cl}=-10E_R$.
  The solid line in each subplot shows the corresponding ground state probability of~\eqref{effHam} and the number of cavity photons
  is $\langle a^\dagger a\rangle=1.3\times 10^{-4}$.}
    \nobreak\medskip
  \end{center}
  \end{figure}

Finally, we investigate the other limit of validity, where a rather
large external classical potential, but only a very low photon
number is given, i.e., a weakly driven cavity. Here the ground state
properties of our model resemble to a very high degree those of the
ordinary Bose-Hubbard model. As mentioned above, an atomic ensemble
interacting with a purely classical potential, has no channels of
dissipation in the absence of spontaneous emission. So unless we
prepare the system in its groundstate, it will show undamped
oscillation. In strong contrast the coupling of the atoms to an even
small intracavity field with a very low photon number opens a
dissipation channel. Although the enhancement of atom number
fluctuations due to fluctuation induced tunneling is small, this
damping still can drive the system into a steady state, very closely
to the adiabatic ground state of~\eqref{effHam}. This is shown in
Fig.~\ref{fig:onsitevar} for the case of two atoms in two wells.
Here we prepare, for different values of on-site interaction, the
atoms in a state perturbed from the ground state of~\eqref{effHam}
with initially no photon in the cavity and a given value of the
classical potential $V_{cl}=-10E_R$. For $g_{1D}=0$, the ground
state is the superfluid state, so Fig.~\ref{fig:onsitevar}(a) is the
generalization of Fig.~\ref{fig:x_av_ph_nr_var} to two atoms.
Although the photon number is only $\langle a^\dagger
a\rangle=1.3\times 10^{-4}$, the system is driven into its ground
state. For increasing interaction strength, the Mott insulator state
becomes more and more favored. Still, the interaction with the tiny
intracavity field enables damping of the atomic evolution towards a
steady state, very close to the adiabatic ground state.

This leads to the conclusion, that, although the cavity field may not lead to
significant modifications of the ground state of the system, the
cavity is a useful tool for faster preparing a system of atoms in its
ground state by opening a dissipation channel, so that it decays towards an eigenstate of the adiabatically eliminated Hamiltonian~\eqref{effHam}.

\section{Atom pumping}\label{sec:atom_pump}
Let us now return to our starting Hamiltonian~\eqref{BHham2} and
consider a second generic model, where the pump laser is not
injected through the cavity mirrors, but directly illuminating the
atoms. This rather small change has a drastic influence on the
physical behavior of this system. In the case of cavity pumping, all
atoms are simultaneously coupled to the same mode. In this way the
cavity field depends on the atomic distribution and long range order
interactions are mediated via the cavity field, influencing the
Mott-insulator to superfluid phase transition. In the new geometry,
only the directly excited atoms coherently scatter photons in the
cavity mode. Due to the position-dependent coupling, the scattered
field amplitude and phase for each atom is strongly position
dependent. Atoms located at nodes are not coupled to the field,
leading to no scattering, whereas atoms at antinodes are maximally
coupled, leading to maximum scattering. Atoms in adjacent wells are
separated by half a wavelength and scatter with opposite phases,
such that their contributions to the scattered field interferes
destructively. Naively one would thus immediately conclude that
atoms forming a state with a homogeneous density scatter no field at
all so that nothing happens~\cite{mekhov1,mekhov2}. Nevertheless,
fluctuations of the density still can allow for some background
scattering which should diminish for lower temperature. For suitable
parameters the corresponding forces start to reorder the atoms
towards a periodic pattern of the atoms, where scattering is
strongly enhanced. This then deepens the optical potential,
stabilizing the pattern in a self-organizing runaway process,
semiclassically described in~\cite{domokos02}.

At $T=0$ quantum fluctuations still can trigger this reorganization.
To study this effect we assume the coherent pump field to form a
broad plane wave propagating transversally to the cavity axis (see
Fig.~\ref{cavity}) replacing cavity pumping. This means that we set
$\eta=0$ and the Hamiltonian~\eqref{BHham2} for constant atom number
$N$ reads as follows:

\begin{eqnarray}\label{BHhamAP}
 H&=&\left( E+J V_{\textrm{cl}}\right)\hat{B}+\hbar\left(U_0J_0N-\Delta_c\right)a^\dagger a+\frac{U}{2}\hat{C}\nonumber
 \\&&+\hbar U_0J a^\dag a\hat{B}+\hbar\eta_{\textrm{eff}}\left(a+a^\dagger\right)\tilde{J}_0\hat{D}.
  \end{eqnarray}

Here we introduced the operator $\hat{D}=\sum_k(-1)^{k+1}\hat{n}_k$
describing the difference in atom number between odd and even sites.
The corresponding Heisenberg equation for the cavity
field~\eqref{eq:heisenberg2} reads as follows:

\begin{equation}\label{eq:heisenberg3}
\dot{a}=\left\{i\left[\De_c-U_0\left(
J_0N+J\hat{B}\right)\right]-\kappa\right\}a-i\eta_{\textrm{eff}}\tilde{J}_0\hat{D}.
\end{equation}

Consequently the Heisenberg equation for the particle operators is:
\begin{multline}\label{eq:heisenberg_particle}
\dot{b}_k =\left(E+JV_{cl}-iU_0Ja^\dagger a\right)\left(b_{k+1} +
b_{k-1}\right)\\-i\eta_{\textrm{eff}}\tilde{J}_0\left(a+a^\dagger\right)(-1)^{k+1}b_k+U\hat{n}_kb_k.
\end{multline}

Hence we see that the occupation number difference drives the cavity
field, which then in turn starts to dephase neighboring atom sites
via the first term in the second line of
Eq.~\eqref{eq:heisenberg_particle}. Note that this interesting part
of the dynamics even survives for deeper lattices when $J$ is
negligibly small and $\tilde{J}_0$ is of order unity. This will
discussed in more detail using various approximations below.

\subsection{Field-eliminated Hamiltonian}\label{sec:field_elim_Ham_AP}

Adiabatic elimination of the field variables is a bit more subtle here as compared to
the cavity pump case discussed before. The scattering amplitude of light into the
cavity mode here depends strongly on the atomic positions. Hence even small
position changes have a large influence on the cavity field dynamics. The maximum photon number
is established when all the atoms are well localized at either only odd or only even lattice sites.
For red atom field detuning this increases the lattice depth and forces the atoms into one of two
stable patterns, where the wells where atoms are located are deeper
than the empty ones. Hence this changes the translational periodicity of the
optical lattice from $\lambda/2$ to $\lambda$. Such bistable behavior was observed by Vuleti\'c and
coworkers~\cite{black03} and explained in a semiclassical treatment~\cite{domokos02}.

Let us now turn to a quantum treatment of atoms and field. Naive
adiabatic elimination encounters a first difficulty, as the
operators $\hat{B},\hat{D}$ do not commute, $[\hat{B},\hat{D}]\neq
0$. Hence this already creates an ordering problem in the formal
steady-state solution of~\eqref{eq:heisenberg3}, which gets even
more difficult when it comes to the replacement of the field
operators to obtain an effective Hamiltonian~\eqref{BHhamAP}.
Unfortunately also the second approach used in the case of cavity
pumping, namely reading off an effective Hamiltonian from the
particle operator Heisenberg equation does not resolve this
problems. Replacing $a$ with the steady-state expression
in~\eqref{eq:heisenberg_particle} leads to a rather complex form, so
that there is no simple way to find a suitable effective Hamiltonian
$H_{\textrm{ad}}$, with $i\hbar\dot{b}_k=[b_k,H_{\textrm{ad}}]$.

Hence we have to resort to the further approximation of neglecting
the term  $\hbar U_0Ja^\dagger a\hat{B}$, compared to $JV_{cl}$.
This still leaves the most important part of the new physics, but
reduces the field equation to the form:

\begin{equation}\label{eq:heisenberg4}
\dot{a}=\left(i\De_c^\prime-\kappa\right)a-i\eta_{\textrm{eff}}\tilde{J}_0\hat{D}.
\end{equation}

The steady-state solution of this equation is immediately at hand and free of ordering ambiguities of non-commuting operators.
\begin{equation}\label{eq:a_SS_AP}
a=\frac{i\eta_{\textrm{eff}}\tilde{J}_0}{
i\Delta_c^\prime-\kappa}\hat{D}.
\end{equation}
%
%
Also the particle operator equation is much simpler within this
approximation:
\begin{multline}\label{eq:heisenberg_particle2}
\dot{b}_k = \left(E+JV_{cl}\right)\left(b_{k-1}+b_{k+1}\right)\\-i\eta_{\textrm{eff}}\tilde{J}_0\left(a+a^\dagger\right)(-1)^{k+1}b_k + U\hat{n}_kb_k.
\end{multline}
In this form one then can find a well defined effective Hamiltonian
only containing particle operators. Let us thus proceed as in
Sec.~\ref{sec:field_elim_Ham} and simply
substitute~\eqref{eq:a_SS_AP} and its adjoint into~\eqref{BHhamAP}.
This yields the effective Hamiltonian:
\begin{equation}\label{eq:effHamAP}
H_{\textrm{ad}}=(E+JV_{\textrm{cl}})\hat{B}+\frac{U}{2}+\frac{\hbar\tilde{J}_0^2\eta_{\textrm{eff}}^2\Delta_c^\prime}{\kappa^2+{\Delta_c^\prime}^2}\hat{D}^2.
\end{equation}

Within first order in $J$ the replacement of the field variables in
the Liouvillean part of the master equation~\eqref{eq:liouvillian}
in this case does not provide an extra terms to be included in the
Hamiltonian. So the effective cavity decay induced dissipation of
the atomic dynamics takes the simple and intuitive form:

\begin{equation}\label{eq:LiouAP}
\mathcal{L}_{\textrm{ad}}\varrho = \frac{\kappa \eta_{\textrm{eff}}^2 \tilde{J}_0^2}{\kappa^2+{\Delta_c^\prime}^2}\left(2\hat{D}\varrho\hat{D}-\hat{D}^2\varrho-\varrho\hat{D}^2\right).
\end{equation}

Note that Eq.~\eqref{eq:effHamAP} with the replacement of the field
operator by its steady-state expression also leads to the same time evolution
as induced by~\eqref{eq:heisenberg_particle2} after symmetrizing with respect to the
field terms. The two approaches thus lead to identical predictions,
which we will exhibit in some more detail in the following.

\subsection{Self-organization of atoms in an optical lattice}\label{sec:QPT_AP}

In this section we investigate the microscopic dynamics of self-ordering near zero temperature
and compare the results of the general model Hamiltonian~\eqref{BHhamAP} and the
corresponding effective Hamiltonian~\eqref{eq:effHamAP}. In order to simplify things,
we keep the approximation from above and neglect $JU_0a^\dagger a$ in the model, i.e.,
 \begin{equation}
\label{BHhamAP2} H=\left( E+J V_{\textrm{cl}}\right)\hat{B}-\hbar\Delta_c^\prime a^\dagger a+\frac{U}{2}\hat{C}+\hbar\eta_{\textrm{eff}}\tilde{J}_0\hat{D}\left(a+a^\dagger\right),
  \end{equation}
with $\Delta_c^\prime=\Delta_c-U_0J_0N$.
Let us point out here, that the Hamiltonian in this approximative form is equivalent to a Hamiltonian describing
1D motion along an optical lattice transverse to the cavity axis. Such a lattice can e.g. be generated
by the pump laser itself as it was studied in~\cite{quantum_seesaw,quantum_seesaw2} to
investigate the onset of the self-organization process~\cite{domokos02,Zippilli04a,black03} at zero temperature.

Similar to that case, the effective Hamiltonian Eq.~\eqref{BHhamAP2} for moderate coupling reproduces
quite well the results of a full Monte-Carlo wavefunction simulation. We have checked this
for a rather small system of two atoms in two wells with periodic boundary conditions. This is the minimal system to study
self-organization but in general sufficient to capture the physics. In this special case the operator $\hat{B}$ simply couples
the ordered $\vert 11\rangle$ state to the state $1/\sqrt{2}\left(\vert20\rangle+\vert02\rangle\right)$, while the operator $\hat{D}^2$ leaves all the basis states $\{\vert 11\rangle,\vert 20\rangle,\vert 02\rangle\}$ unchanged. It simply
leads to a relative energy shift. Hence starting from a perfectly ordered atomic state (the analog of the Mott insulator state) the Hamiltonian part of the time evolution of the system couples it to the symmetric superposition of ordered states. In an adiabatic limit those ordered states are correlated with a coherent field $\pm \alpha$ in the cavity. Thus without damping the evolution would simply read:
\begin{equation}
\vert \psi(t)\rangle = \cos\left(2\omega t\right)\vert 11,0\rangle + \frac{i\sin\left(2\omega t\right)}{\sqrt{2}}\left(\vert 20,2\alpha\rangle + \vert 02,-2\alpha\rangle\right).
\end{equation}
where the frequency $\omega$ is given by the $(E+J
V_{\textrm{cl}})/\hbar$. Here $\vert 11,0\rangle$ is the state with
one atom in each well and zero photons, whereas $\vert
20,\alpha\rangle$ ($\vert 02,-\alpha\rangle$) corresponds to the
state with both atoms in the left (right) well, and the cavity field
being in a coherent state with amplitude $2\alpha$ ($-2\alpha$). The
factor 2 is due to constructive interference of the fields,
scattered by the ordered atoms. In the Mott state, the scattering
fields cancel each other.

Note that such an entangled superposition of different atomic states
and fields cannot be reproduced by any classical or mean field
evolution and requires a genuine quantum description. If on-site
interaction is added the amplitude of this oscillations decreases
due to extra relative different phase changes of the self-ordered
and the Mott state.

Of course we now have to add the effect of dissipation via cavity
loss. We will see that even single cavity photon decay events
strongly perturb the system evolution. This can be immediately seen
by applying the photon annihilation operator to the entangled
atom-field state, i.e.,

\begin{equation}\label{eq:coh_states}
\vert \psi(t)^\prime\rangle\propto a \vert \psi (t)\rangle \propto  \vert 20,2\alpha\rangle - \vert 02,-2\alpha\rangle.
\end{equation}

This procedure  projects out the Mott contributions to the state as
they are connected to zero photons. Surprisingly in addition it also
blocks further tunneling by introducing a minus sign between the two
ordered states. At this point coherent atomic time evolution stops
until a second photon escapes and re-establishes the plus sign. This
then allows tunnel coupling back to the Mott insulator state again.
In this sense self-ordering is an instantaneous projective process
here, where the cavity acts as measurement apparatus asking a sort
of yes/no ordering question.

\begin{figure}[htp]
    \begin{center}
\includegraphics[width=0.8\hsize]{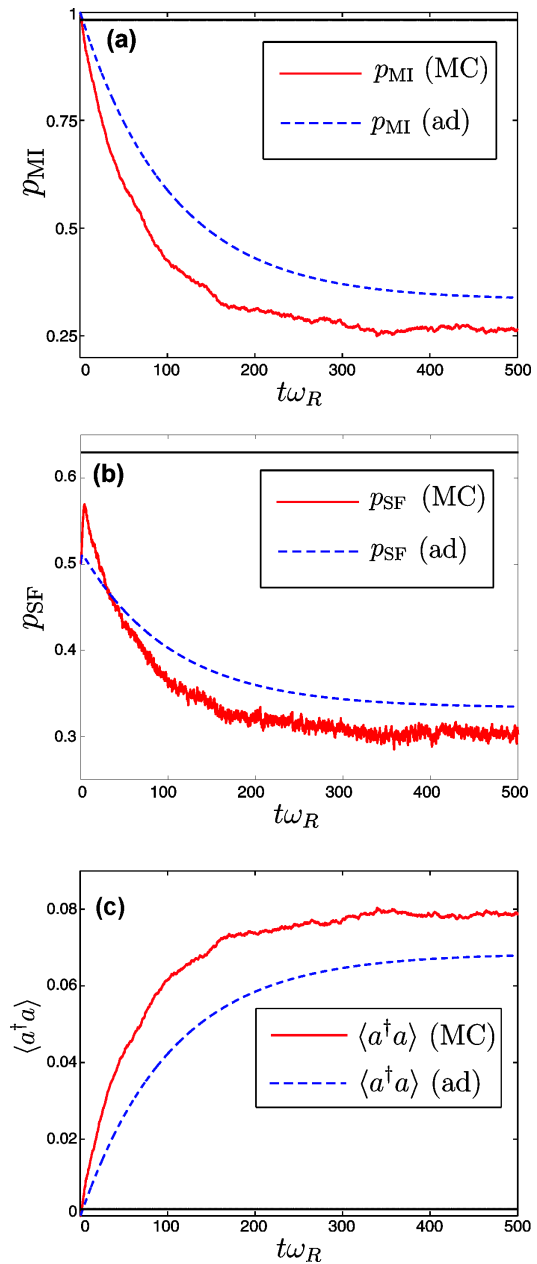}
  \caption{\label{fig:MCsim:AP}(color online) (a) Contribution of the Mott insulator state $p_{\textrm{MI}}$ in a system of two atoms in two wells. The solid line shows the results of a Monte-Carlo simulation, corresponding to~\eqref{BHhamAP2} with dissipation via cavity loss. The dashed line depicts the solution of a master equation with effective Hamiltonian~\eqref{eq:effHamAP} and
Liouvillean~\eqref{eq:LiouAP}. The constant line, shows the ground state value of this contribution of the effective Hamiltonian. (b) The same for the contribution of the superfluid state $p_{\textrm{SF}}$. (c) shows analogue results for the mean photon number. Parameters are  $V_{\textrm{cl}}=-10E_R,\, (\kappa,U_0,\eta_{\textrm{eff}})=(4,-0.1,1)\,\omega_R$ and $\Delta_c=U_0J_0N+\kappa$. }
   \nobreak\medskip
   \end{center}
  \end{figure}

The fact, that for transverse pumping the adiabatic field state
associated with the Mott insulator is an intracavity vacuum
decouples this state from further dynamics even in the presence of
dissipation.  This creates numerical difficulties and prohibits an
approximation of the dissipative dynamics by the adiabatic ground
state values of~\eqref{eq:effHamAP} only. As soon as a photon leaks
out of the cavity, the contribution of the Mott-insulator state is
canceled, no matter how large it, corresponding to a given on-site
interaction, might be. Hence, every initial state evolves into a
superposition of the ordered states and the ground state values of
the effective Hamiltonian do not make much sense. Nevertheless,
including the damping via the effective
Liouvillean~\eqref{eq:LiouAP} approximately reveals the complete
dynamics. In Fig.~\ref{fig:MCsim:AP} we show the results of a
Monte-Carlo simulation of the dynamics of the Mott and the
superfluid contribution, corresponding to~\eqref{BHhamAP2} and
compare it with a solution of the master equation, consisting of the
Hamiltonian~\eqref{eq:effHamAP} and Liouvillean~\eqref{eq:LiouAP},
where the field variables are eliminated. Furthermore, the
restriction of the Hilbert space to the two states
of~\eqref{eq:coh_states} and $\vert 11,0\rangle$, allows for a proof
of the accuracy of our assumption, concerning the fast evolution of
the cavity field. We use the coefficients $c(t),\tilde{c}(t)$
(calculated with the Monte Carlo simulation) of $\vert
\psi(t)\rangle = \tilde{c}(t)\vert 11,0\rangle + c(t)(\vert
20,2\alpha\rangle\pm\vert 02,-2\alpha\rangle)$ to construct a purely
atomic state $\vert \varphi(t)\rangle = \tilde{c}(t)\vert 11\rangle
+ c(t)(\vert 20\rangle\pm\vert 02\rangle)$. Then the mean photon
number, calculated with the effective photon
operator~\eqref{eq:a_SS_AP} agrees very well with the real mean
photon number, i.e.,
\begin{equation}
\frac{\eta_{\textrm{eff}}^2\tilde{J}_0^2}{{\Delta_c^\prime}^2+\kappa^2}\langle
\varphi(t)\vert \hat{D}^2\vert \varphi(t)\rangle \approx
\langle\psi(t)\vert a^\dagger a\vert\psi(t)\rangle.
\end{equation}


%

\section{Conclusions}\label{sec:conclusions}

Based on an approximative Bose-Hubbard type model descriptions, we
have shown that quantum characteristics of light fields generating
optical potentials lead to shifts in quantum phase transition points
and play a decisive role in the microscopic dynamics of the
transition process. While many physical aspects can be already
captured by effective Hamiltonians with rescaled parameters, cavity
mediated long-range interactions also play an important role and add
a new nonlocal element to optical lattices dynamics for atoms. In
that context even small modifications in the setup, from cavity pump
to transverse pump, have a drastic influence on the behavior of the
system on a microscopic level. We have seen that the Bose-Hubbard
Hamiltonian for the former system can, in a certain parameter
regime, be significantly simplified by adiabatically eliminating the
field variables. Although the cavity has influence on its shape, the
Mott insulator to superfluid phase transition occurs similar to
classical optical lattices. For transverse pumping this is not the
case. Here, the fields scattered by the atoms in the uniform Mott
state cancel and completely suppress scattering. In parallel new
ordered states with maximal coupling of pump and cavity field appear
and the dynamics favors a superposition of these two ordered states
correlated with coherent field states with phase difference $\pi$.
Hence the dynamics generates strong atom field entanglement and
large effective optical nonlinearities even in the limit of linear
weak field scattering.

Of course the various approximations used to derive our effective
Hamiltonians still leave a lot of room for improvements and we could
only touch a very small part of the physical effects and
possibilities contained in these model. Fortunately the experimental
progress in this field is spectacularly fast and several groups now
have set up optical lattices with cavity
fields~\cite{Esslinger,Reichel07,Brennecke07,Slama07,Gupta07} and
intriguing potential applications of such systems were already
proposed~\cite{Meiser07}, so that one can expect a fast and exciting
further development of this field.

\section*{Acknowledgments}
The authors would like to thank M. Lewenstein, G. Morigi, S. Fern\'andez-Vidal, A.
Micheli, and A. Vukics for useful discussions. This work was funded by the Austrian Science Fund (P17709 and S1512).
After completion of this work we became aware of related parallel work by M. Lewenstein and coworkers, which treats many aspects of this model in the thermodynamic limit~\cite{Larson07}.

\newpage

\begin{thebibliography}{99}
\bibitem{comment1}It is also possible, first to eliminate the excited state in the single particle dynamics, which leads to the same expression~\eqref{eq:adhamSP} and then implement this expression in second quantization formalism.
\end{thebibliography}
\end{document}